\documentclass[useAMS]{mn2e}

\usepackage{amssymb,amsmath,psfig,times}
\def\gsim{ \lower .75ex \hbox{$\sim$} \llap{\raise .27ex \hbox{$>$}} }
\def\lsim{ \lower .75ex\hbox{$\sim$} \llap{\raise .27ex \hbox{$<$}} }

\def\sc{Schwarzschild}

\title[Canonical high power blazars]
{Canonical high power blazars}  
\author[G. Ghisellini \& F. Tavecchio]
{G. Ghisellini\thanks{Email:
gabriele.ghisellini@brera.inaf.it} and F. Tavecchio
\\
INAF -- Osservatorio Astronomico di Brera, Via Bianchi 46, I--23807 Merate, Italy\\
}

\begin{document}  

\maketitle

\begin{abstract}
The jets of powerful blazars propagate within regions relatively 
dense of radiation produced externally to the jet.
This radiation is a key ingredient to understand the origin of the high 
energy emission of blazars, from the X--ray to the $\gamma$--ray energy band.
The main components contributing to the external radiation field are the 
accretion disk emission, including its X--ray corona, the broad line region, 
the infrared emitting torus and the cosmic background radiation.
Their importance changes as a function of the distance from the black 
hole and of the value of the bulk Lorentz factor of the jet.
These external radiation fields control the amount of the 
inverse Compton radiation with respect to the synchrotron flux.
Therefore the predicted spectral energy distribution (SED)
will depend on where the jet dissipates part of its energy to 
produce the observed radiation.
We investigate in detail how the SED changes as a function of the 
location of the jet dissipation region, by assuming rather 
``standard" (i.e. ``canonical'') prescriptions for the accretion 
disk and its X--ray corona, the profile of the jet magnetic field 
and the external radiation.
We confirm that most of the dissipation, if producing the $\gamma$--ray 
flux we see, must occur at hundreds of \sc\ radii from the black hole, 
to avoid the $\gamma$--$\gamma\to e^\pm$ process, and the consequent 
re--emission by the produced pairs.
The magnetic energy density of a ``canonical" jet almost never dominates 
the radiative cooling of the emitting electrons, and consequently
the inverse Compton flux almost always dominates the bolometric output.
This is more so for large black hole masses.
Dissipation taking place beyond the broad line region is particularly
interesting, since it accounts in a simple way for the largest inverse 
Compton to synchrotron flux ratios accompanied by an extremely hard 
X--ray spectrum. 
Furthermore it makes the high power blazars at high redshift
useful tools to study the optical to UV cosmic backgrounds.
\end{abstract}


\section{Introduction}

Relativistic jets in blazars transport energy in the form of bulk motion of protons,
leptons and magnetic field.
When part of this power is dissipated, the particles emit the beamed radiation we
observe, consisting of two broad humps.
The origin of the low frequency hump is well established, believed to be
synchrotron radiation from relativistic (in the comoving frame) leptons.
The nature of the high energy hump is a controversial issue. 
In general, we can group the existing models into three families:
i) the high energy radiation is generated by the same leptons producing the
synchrotron, through the inverse Compton process 
(Maraschi, Ghisellini \& Celotti 1992;
Dermer \& Schlickeiser 1993; 
Sikora, Begelman \& Rees 1994; 
Ghisellini \& Madau 1996; 
Bloom \& Marscher 1996;
Celotti \& Ghisellini 2008);
ii) There are two populations of leptons, one accelerated directly by the 
acceleration mechanism (i.e. shocks), and a second one resulting from 
cascades initiated by ultra--relativistic protons 
(Mannheim 1993;
M\"ucke et al. 2003; 
B\"ottcher 2007);
iii) Ultrarelativistic protons emit by the proton--synchrotron process at
high energies (Aharonian 2000; M\"ucke \& Protheroe 2001).

Another issue of debate is the role of electron--positron pairs. 
Sikora \& Madejski (2000) discussed this problem concluding that,
though they can exist, their amount is limited to a few pairs per proton,
and a similar conclusion was reached by Celotti \& Ghisellini (2008) 
analysing a large number spectral energy distributions (SEDs) for deriving
the jet powers of blazars.
This contrasts the idea, originally put forward by Blandford \& Levinson (1995)
that the $\gamma$--rays spectrum is the superposition of the spectra
originating at different distances from the black hole, each one cutted--off
at a different $\gamma$--ray energy as a result of photon--photon absorption.
In this scheme a large amount of pairs is produced, since the high energy hump carries
most of the emitted power: if this is partly absorbed, we transform most of the 
total power into (energetic) pairs.

Ghisellini \& Madau (1996) argued that the Blandford \& Levinson idea 
has one important observational consequence in powerful blazars
with a standard accretion disk: since the pairs are born relativistic,
they contribute to the emission mainly by inverse Compton scattering 
the dense UV radiation field coming from the accretion disk.
In fact, if we want the emission region be compact, it is reasonable to locate
it close to the accretion disk and its X--ray corona.
This implies that the coronal X--rays are targets for the 
$\gamma$--$\gamma \to e^\pm$ process, and that the accretion disk UV photons
become seeds for the scattering made by the newly born pairs.
The resulting radiation is mainly in the X--ray band, that should have a 
power comparable to the power absorbed in the $\gamma$--ray band, contrary
to what observed.
This led Ghisellini \& Madau (1996) to conclude that the dissipation region
in blazar jets cannot be very close to the accretion disk.
On the other hand the observed fast variability argues for not too large distances.
Taken together, these two limits strongly suggest that there is a preferred distance
at which {\it the $\gamma$--ray radiation is produced}, at some hundreds of \sc\ radii.
Since this argument makes use of the reprocessed radiation produced by the pairs,
it cannot be applied when most of the radiation is produced below the pair 
production energy threshold and/or if the accretion disk is
radiatively inefficient: in both cases very few pairs are created.
Indeed, Katarzynski \& Ghisellini (2007) proposed a jet model in which
a dissipation close to the accretion disk resulted in a low level 
$\gamma$--ray emission, with no contradiction with the existing X--ray data.

These arguments, although correct, are qualitative, and 
in our opinion a detailed computation of the reprocessing due to pairs
is not yet present in the literature.  
Therefore one of the aims of the present paper is to derive
some limits on the location of the dissipation region in 
the jet of powerful blazars.
We will do it in the framework of the ``leptonic" class of models 
[family i) mentioned above], and we limit our analysis to blazars
having a ``standard" (Shakura \& Sunyaev 1973) accretion disk.
This implies to consider Flat Spectrum Radio Quasars (FSRQs, and their likely
parent population, FR II radio--galaxies) and not BL Lac objects, that are
likely to have radiatively inefficient accretion flows (see 
Celotti \& Ghisellini 2008 and Ghisellini \& Tavecchio 2008 for more discussion
about this point).

While doing this, we will study the relative importance of 
different sources of seed photons as a function of the distance $R_{\rm diss}$
of the dissipation region from the black hole, and their spectrum 
as seen in the comoving frame.
This is the second aim of the paper. 
Besides the radiation coming directly from the accretion disk and
its X--ray corona, we will consider the radiation produced in the 
Broad Line Region (BLR) and in a relatively more distant dusty torus, intercepting
a fraction of the disk radiation and re--emitting it in the infrared.
Finally, we also include the contribution of the Cosmic Microwave Background
(CMB), important for very large (beyond 1 kpc) $R_{\rm diss}$.

The term ``canonical high power blazars'' refers to the rather standard
choice for both the environment of these sources and their jets:
\begin{enumerate}
\item 
The accretion disk is a standard, ``Shakura \& Syunyaev" (1973) disk.
\item
Above this accretion disk, there is an X--ray 
corona, emitting a luminosity less than, but comparable to, the luminosity
emitted by the accretion disk.
\item 
the BLR and the IR torus are located at distances
that scale as the square root of the disk luminosity
(Bentz et al. 2006;
Kaspi et al. 2007;
Bentz et al. 2008;
see the discussion about this point made in Ghisellini \& Tavecchio 2008).

\item 
The jet is assumed to dissipate only a fraction of its 
total power, which is then conserved. 
After the acceleration phase, possibly magnetic in origin, 
we assume that also the Poynting flux
is conserved (see e.g. Celotti \& Ghisellini 2008).
\end{enumerate}
Our study is not completely new, since several of its ``ingredients"
have already been discussed in the literature.
The paper more german to our is Dermer et al. (2009), but we include 
some new ingredients.
The novel features of our investigation concern mainly: i) the inclusion of the
X--ray corona as an important producer  of target photons for the 
$\gamma$--$\gamma \to e^\pm$ process; ii) the calculation of the emitting
particle distribution, including pair creation; 
iii) the effects of jet acceleration at small distances from the black hole;
iv) the strict link between the properties of the accretion disk and the 
amount of
the external radiation and v) the overall scenario allowing to
describe in a more general way (than done before) 
the SED properties of high power jets at all scales.

The paper is divided into four parts.
In the first we study the different sources of external radiation,
and the corresponding energy densities, as seen in the comoving frame,
as a function of distance from the black hole and as a function of the
bulk Lorentz factor of the jet. 
In the second part of the paper we investigate the role of pair production 
processes when the dissipation region is close to the black hole, with the aim
to find quantitative constraints.
In the third part we construct the expected SED as a function of distance,
highlightening what are the relevant external seed photons.
Finally, in the fourth part, we apply some of the above results
and considerations when modelling the SED of some blazars, used
as illustrative examples, and we check if our scenario can
reproduce the phenomenological blazar sequence (Fossati et al. 1998).

We use a cosmology with $h_0=\Omega_\Lambda=0.7$ and $\Omega_{\rm M}=0.3$.
We also use the notation $Q=10^xQ_X$ in cgs units, unless noted otherwise.

\section{Setup of the model}

Our model is characterised by the following setup.
The accretion disk extends from $R_{\rm in}=3 R_{\rm S}$  to 
$R_{\rm out}=500 R_{\rm S}$ ($R_{\rm S}$ is the \sc\ radius) 
and is producing a total luminosity
$L_{\rm d}=\eta \dot M c^2$, where $\dot M$ is the accretion rate and $\eta$
is the accretion efficiency.
Locally, its emission is black--body, with a temperature
\begin{equation}
T^4 \, =\, {  3 R_{\rm S}  L_{\rm d }  \over 16 \pi\eta\sigma_{\rm MB} R^3 }  
\left[ 1- \left( {3 R_{\rm S} \over  R}\right)^{1/2} \right]   
\end{equation}
Below and above the accretion disk there is a hot corona, emitting UV  
and X--rays with a luminosity $L_X=f_X L_{\rm d}$.
For simplicity, the corona is assumed to be homogeneous between 3 and 30 \sc\ radii.
The spectrum is assumed to be a cut--off power law: 
$L_X(\nu)\propto \nu^{-\alpha_X}\exp(-\nu/\nu_{\rm c})$.

The broad line region (BLR) is assumed to be a shell located at a distance
\begin{equation}
R_{\rm BLR} \, =\,  10^{17} \,  L_{\rm d, 45}^{1/2}\,\, {\rm cm}
\end{equation}
It reprocesses a fraction $f_{\rm BLR}$ of $L_{\rm d}$ in lines, especially
the hydrogen Lyman--$\alpha$ line, and continuum.
Following Tavecchio \& Ghisellini (2008), we assume that 
the spectral shape of the BLR observed in the comoving frame is a black--body 
peaking at a factor $\Gamma$ times the (rest frame) frequency of the Lyman--$\alpha$ line.

We also assume the presence of a torus (see B{\l}azejowski et al. 2000; 
Sikora et al. 2002), at a distance
\begin{equation}
R_{\rm IR} \, =\,  2.5\times 10^{18} \,  L_{\rm d, 45}^{1/2}\,\, {\rm cm}
\end{equation}
reprocessing a fraction $f_{\rm IR}$ of the disk radiation in the infrared.
Note that both $R_{\rm BLR}$ and $R_{\rm IR}$ scale as the square root
of $L_{\rm d}$: this implies that, in the lab frame, the radiation energy densities
of these two components are constant, as long as $R_{\rm diss}$ is smaller than
these two radii.
We emphasise that our treatment of the torus emission is approximate:
it is likely that the torus itself is a complex structure, possibly clumpy 
(Nenkova et al. 2008) with a range of radii, extending also quite close to
the black hole, where the temperature is just below dust sublimation 
(i.e. $\sim 1500$ K).
Our approach follows in part the results of Cleary et al. (2007),
finding weak signs of hot dust emission in the studied spectra, and
partly is dictated by simplicity.

The emitting region is moving with a velocity $\beta c$ corresponding to a bulk
Lorentz factor $\Gamma$.
We call $R_{\rm diss}$ the distance of the  dissipation region from the black hole.
We consider either a constant $\Gamma$ or include an acceleration phase of the kind
(see e.g. Komissarov et al. 2007; Vlahakis \& K\"onigl 2004):
\begin{equation}
\Gamma \, =\,  \min\left[\Gamma_{\rm max}, 
\left({R\over 3R_{\rm S}}\right)^{1/2}\right]
\end{equation}

When the acceleration phase is taken into account, the jet is assumed
to be parabolic in shape, becoming conical when $\Gamma$ reaches
its maximum value (see e.g. Vlahakis \& K\"onigl 2004).
Calling $r$ the cross sectional radius of the jet, and $R$ the distance
from the black hole, we have
\begin{eqnarray}
r \, &=&\,  \phi\, R^{1/2}, \quad \Gamma \le \Gamma_{\rm max}  \nonumber \\
r \, &=&\, \psi \, R, \quad\quad\, \Gamma \ge \Gamma_{\rm max}
\end{eqnarray}
where $\psi$ is the semi--aperture angle of the jet in its conical part,
and the constant $\phi$ is fixed by assuming that at the start of the jet
we have $r_o=R_0=3R_{\rm S}$. 
The parabolic and conical parts of the jet connect at 
$R=3R_{\rm S}\Gamma^2_{\rm max}$.

The total power carried by the jet, $P_{\rm j}$, was assumed
in Ghisellini \& Tavecchio (2008) to be related to 
the mass accretion rate, i.e.  $P_{\rm j}=\eta_{\rm j} \dot M c^2$.
Since Celotti \& Ghisellini (2008) 
found that $P_{\rm j}$ is greater than
the disk accretion luminosity, i.e. $P_{\rm j}\gsim L_{\rm d}$,
we assumed (in Ghisellini \& Tavecchio 2008)
that $\eta_{\rm j}$ was greater than the
corresponding efficiency of transforming the accretion 
rate $\dot M$ in disk luminosity.
When fitting the data of specific sources,
we do not specify a priori the total jet luminosity,
which is instead a result of the modelling 
(once we assume how many protons there are for each emitting
lepton).
Therefore $P_{\rm j}$ is not an input parameter in our scheme, 
but it is a quantity derived a posteriori.
What we will specify, instead, is the power injected
in the dissipation region, that is related to the power carried
by the jet in relativistic electrons (and by cold protons,
once we specify how many protons there are per emitting electron). 

We assume a value of the magnetic field in the dissipation region,
that corresponds to a Poynting flux $P_{\rm B}$. 
If the jet is magnetically accelerated, then the initial (i.e. close to $R_0$)
$P_{\rm B}$ should be of the same order of $P_{\rm j}$, becoming
$P_{\rm B}=\epsilon_{\rm B}P_{\rm j}=$const when $\Gamma$ reaches its maximum value.
To describe the profile of $P_{\rm B}$ we assume the
following prescription:
\begin{equation}
P_{\rm B} \, =\, \pi r^2 \Gamma^2 c U_{\rm B} \,
=\,  P_{\rm j}  \left[ 1- {\Gamma\beta \over \Gamma_{\rm max}\beta_{\rm max} }
\left(1-\ \epsilon_{\rm B}\right)\right]
\end{equation}
In this way $P_{\rm B}=P_{\rm j}$ initially, becoming a 
constant fraction
$\epsilon_{\rm B}$ of $P_{\rm j}$ when the jet is conical.
Here $U_{\rm B}=B^2/(8\pi)$ is the magnetic energy density.

The energy distribution of the particles responsible for the emission
is derived through the continuity equation, assuming 
a continuous injection of particles throughout the source lasting
for a finite time. This time is the light crossing time 
$t_{\rm cross}=r_{\rm diss}/c$, where $r_{\rm diss}$ is the
size of the emitting blob, located at the distance $R_{\rm diss}$ from 
black hole.
We always calculate the particle distribution at this time.
The reason for this approach is suggested by the fast variability
shown by blazars, indicating that the release of energy 
is short and intermittent.
Besides, we believe that this approach is the simplest that can
nevertheless describe in some detail the particle distribution.
In fact, it allows to neglect: i) adiabatic losses
(important after $r_{\rm diss}/c$, which is also the time needed 
to double the radius);
ii) particle escape (again important for times longer than $r_{\rm diss}/c$) and
iii) the changed conditions in the emitting region (since the
source is travelling and expanding, the magnetic field changes).

High energy particles can radiatively cool in a time shorter than
$t_{\rm cross}$. Let us call $\gamma_{\rm cool}$ the energy
of those particles halving their energy in a time $t_{\rm cross}$.
Above $\gamma_{\rm cool}$, and at $t=t_{\rm cross}$, 
the particle energy distribution $N(\gamma, t_{\rm cross})$
can  be found by solving
%
%
%
\begin{equation}
{\partial \over \partial \gamma }\left[ \dot \gamma N(\gamma, t_{\rm cross}) \right] + Q(\gamma)
+P(\gamma)\, =\, 0
\label{cont}
\end{equation}
where $\dot\gamma$ is the
cooling rate of a particle of energy $\gamma mc^2$, $Q(\gamma)$ is the source
term (i.e. the injection of primary particles) assumed constant in time, and
$P(\gamma)$ is the term corresponding to the electron--positron pairs that 
are produced in photon--photon collisions.
The formal solution of Eq. \ref{cont} is
\begin{equation}
N(\gamma)  \, =\, { \int_{\gamma}^{\gamma_{\rm max}} [Q(\gamma)+P(\gamma)] d\gamma
\over \dot \gamma }, \quad \gamma>\gamma_{\rm cool}
\label{ngamma}
\end{equation}
When electrons with $1<\gamma<\gamma_{\rm cool}$ do not cool in $t_{\rm cross}$
we approximate the low energy part of $N(\gamma)$ with 
\begin{equation}
N(\gamma)  \, \sim \, t_{\rm cross} [Q(\gamma)+P(\gamma)], 
\quad \gamma<\gamma_{\rm cool}
\label{ngamma2}
\end{equation}
%
%
Note that, within our assumptions, the particle distribution of
Eq. \ref{ngamma} and Eq. \ref{ngamma2}
correspond to the maximum $N(\gamma)$. 
The injection of primary particles $Q(\gamma)$ is a smoothly joining
broken power law:
\begin{equation}
Q(\gamma)  \, = \, Q_0\, { (\gamma/\gamma_{\rm b})^{-s_1} \over 1+
(\gamma/\gamma_{\rm b})^{-s_1+s_2} }
\label{qgamma}
\end{equation}
where $\gamma_{\rm b}$ is a break energy. 
The pair injection term $P(\gamma)$ corresponds to the $\gamma$--$\gamma \to e^\pm$
process only, and it is calculated with 
the prescriptions given by Svensson (1987) and Ghisellini (1989).

The total power injected in the form of relativistic electrons,
calculated in the comoving frame, is
\begin{equation}
P^\prime_{\rm i} \, = \, m_{\rm e}c^2 V
\int_1^{\gamma_{\rm max}} \gamma Q(\gamma)d\gamma
\label{leprime}
\end{equation}
where $V=(4\pi/3)r_{\rm diss}^3$ is the emitting volume.
Note that this is not equivalent to the power that the jet transports
in the form of relativistic particles (as measured in the comoving frame),
since $P^\prime_{\rm i}$ includes also the energy that will be 
emitted, and possibly transformed into pairs.

\section{Energy densities}

\begin{figure}
\psfig{figure=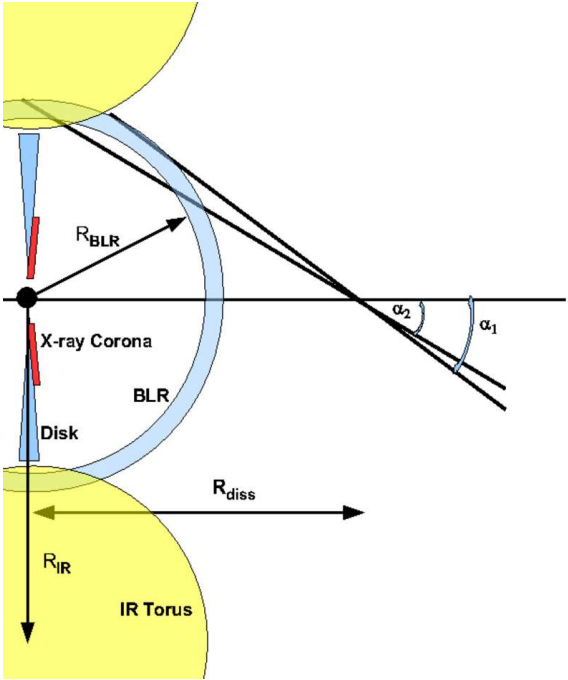,width=9cm,height=9cm}
\caption{Cartoon illustrating the accretion disk, its X--ray corona,
the broad line region and
a schematic representation of the IR torus.
At the distance $R_{\rm diss}$ the jet is assumed to dissipate.
At this distance, here assumed to be outside the BLR, we
label the relevant angles for calculating the contribution of the BLR
radiation to the corresponding energy density.
}
\label{cartoon}
\end{figure}

\subsection{Direct disk radiation}

Each annulus of the accretion disk
is characterised by a different temperature and it is seen under 
a different angle $\xi$ (with respect to the jet axis),
thus its radiation is boosted in a different way.
A stationary observer with respect to the black hole (lab frame)
will see a flux, integrated over all annuli, given by
\begin{equation}
F_{\rm d}(\nu) \,=\, 2\pi \int^1_{\mu_{\rm d}} I(\nu)d\mu \, = \,
2\pi\int^1_{\mu_{\rm d}}  { 2h\nu^3/c^2 \over \exp[ h\nu /(kT)] -1}  d\mu
\end{equation}
where $\mu=\cos\xi$, and $\mu_{\rm d}$ is given by 
\begin{equation}
\mu_{\rm d} \, =\, [1+R^2_{\rm out} /R^2_{\rm diss} ]^{-1/2}
\end{equation}
In the comoving frame of the blob, frequency are transformed as:
\begin{equation}
\nu^\prime =b\nu, \qquad  b\, \equiv\, \, \Gamma(1-\beta\mu)
\end{equation}
and solid angles transform as:
\begin{equation}
d\Omega^\prime\, =\, {d\Omega \over b^2}\, =\, 2\pi {d\mu \over b^2}
\end{equation}
where primed quantities are in the comoving frame.
The intensities as seen in the comoving frame transform as:
\begin{equation}
I^\prime_{\rm d} (\nu^\prime) \, =\, b^3 I_{\rm d}(\nu) \, =\, b^3 I_{\rm d}(\nu^\prime/b)
\end{equation}
The specific radiation energy density 
seen in the comoving frame is
\begin{equation}
U^\prime_{\rm d}(\nu^\prime) \, =\, {1\over c} 
\int  I^\prime_{\rm d} (\nu^\prime) d\Omega^\prime \, =
{2\pi\over c} 
\int^1_{{\mu_{\rm d}}}  {I^\prime_{\rm d}(\nu^\prime) \over b^2}  d\mu  
\end{equation}

\subsection{Radiation from the X--ray corona}

According to our assumptions, the
total radiation energy density $U^\prime_X$ of this component is
(see e.g. Ghisellini \& Madau 1996):
\begin{eqnarray}
U^\prime_X  &=&  
{f_X L_{\rm d} \Gamma^2 \over \pi R_X^2 c}  
\left[ 1-\mu_X-\beta(1-\mu_X^2)+{\beta^2\over 3}(1-\mu_X^3) \right] \nonumber \\
\mu_X &=&  [1+R^2_X/R^2_{\rm diss}]^{-1/2} 
\end{eqnarray}
where $R_X$ is the extension of the X--ray corona.

\subsection{BLR radiation}

Within $R_{\rm BLR}$ the corresponding energy density seen in the comoving frame
can be approximated as (Ghisellini \& Madau 1996):
\begin{equation}
U^\prime_{\rm BLR} \, \sim \,  { 17 \Gamma^2 \over 12} \, { f_{\rm BLR} L_{\rm d}
\over 4\pi R_{\rm BLR}^2 c} 
\,\,  \quad R_{\rm diss}< R_{\rm BLR}
\label{ublr1}  
\end{equation}
At distances much larger than $R_{\rm BLR}$, and calling $\mu=\cos\alpha$, 
we have 
(see the cartoon in Fig. \ref{cartoon})
\begin{eqnarray}
U^\prime_{\rm BLR} \, &\sim& \,  \, 
{ f_{\rm BLR} L_{\rm d} \over 4\pi R_{\rm BLR}^2 c} \, {\Gamma^2 \over 3\beta} 
 \, [ 2(1-\beta\mu_1)^3- (1-\beta\mu_2)^3  \nonumber \\
 &-&(1-\beta)^3)]
\,\,  \qquad R_{\rm diss}\gg R_{\rm BLR} \nonumber \\
\mu_1 \, &=&\, [1+R^2_{\rm BLR}/R^2_{\rm diss}]^{-1/2}  \nonumber \\
\mu_2 \, &=&\, [1 - R^2_{\rm BLR}/R^2_{\rm diss}]^{1/2} 
\label{ublr2}  
\end{eqnarray}
For $R_{\rm diss} \gsim R_{\rm BLR}$ the exact value of $U^\prime_{\rm BLR}$
depends on the width of the BLR, which is poorly known.
For this reason, in the range $R_{\rm BLR}<R_{\rm diss}<3R_{\rm BLR}$ we
use a simple (power--law) interpolation.

The BLR is assumed to ``reflect" (Compton scatter) a fraction $f_{\rm BLR,X}$ 
(of the order of 1 per cent) of the corona emission.
The existence of this diffuse X--ray radiation is a natural outcome 
of photo--ionisation  models for the BLR (see e.g. Tavecchio \& Ghisellini 2008, 
Tavecchio \& Mazin 2009). 
The assumed value, $f_{BLR, X}\sim 0.01$,  
is the average value found for typical parameters of the clouds.

\subsection{Radiation from the IR torus}

This component scales as $U_{\rm BLR}$, but substituting $R_{\rm BLR}$ 
with  $R_{\rm IR}$. We have 
\begin{equation}
U^\prime_{\rm IR} \, \sim  \, { f_{\rm IR} L_{\rm d}\, \Gamma^2 
\over 4\pi R_{\rm IR}^2 c} 
\,\, \quad R_{\rm diss}< R_{\rm IR}
\end{equation}
For $R_{\rm diss}>R_{\rm IR}$ we have the same behaviour as in Eq. \ref{ublr2},
but with $R_{\rm IR}$ replacing $R_{\rm BLR}$.
In a $\nu F_\nu$ plot the (lab frame) peak frequency of this component is assumed
to be at $\nu_{\rm IR} = 3\times 10^{13}$ Hz (see Cleary et al. 2007),
independent of the disk luminosity, since $R_{\rm IR}$ scales as $L_{\rm d}^{1/2}$.
The corresponding temperature is $T_{\rm IR} = h\nu_{\rm IR}/(3.93 k)$ 
(we must use the factor 3.93, instead of the usual 2.82, because we are
using the peak frequency in $\nu F_\nu$).
In the comoving frame this corresponds to
\begin{equation}
T^\prime_{\rm IR} \, \sim\,  370\, b\, \,\, {\rm K}  
\end{equation}

\subsection{Radiation from the host galaxy bulge}

The bulge of the galaxy hosting the blazar can be a non--negligible
emitter of ambient optical radiation (see e.g. 
(Stawarz, Sikora \& Ostrowski 2003).
Within the bulge radius $R_{\rm star}$ emitting a luminosity $L_{\rm star}$
we have
\begin{equation}
U^\prime_{\rm star} \, = \, \Gamma^2 \, {L_{\rm star} \over 4\pi R_{\rm star}^2 c} 
\end{equation}
As an order of magnitude estimate, we have
$U^\prime_{\rm star} \sim 10^{-10} \Gamma^2 $ erg cm$^{-3}$
using  $L_{\rm star} = 3 \times 10^{44}$ erg s$^{-1}$ 
produced within a bulge radius $R_{\rm star}\sim $1 kpc.
When $R_{\rm diss}>R_{\rm star}$, $U^\prime_{\rm star}$ decreases
in an analogous way as $U^\prime_{\rm IR}$ and $U^\prime_{\rm BLR}$
(once we substitute $R_{\rm star}$ to $R_{\rm IR}$ or $R_{\rm BLR}$).

\subsection{Radiation from the cosmic background}

The energy density of the Cosmic Background Radiation (CMB),
as seen in the comoving frame, is

\begin{equation}
U^\prime_{\rm CMB} \, =\, a T_0^4 \Gamma^2 (1+z)^4
\end{equation}
where $a=7.65\times 10^{-15}$ erg cm$^{-3}$ deg$^{-4}$ and
$T_0=2.7$ K is the temperature of the CMB now (i.e. at $z=0$).

\subsection{Magnetic field}

Following our prescriptions concerning the power carried in the form
of Poynting flux, we have
\begin{equation}
U^\prime_{\rm B} \, = \, {P_{\rm B} \over \pi r^2_{\rm diss} \Gamma^2 c } \, =\, 
{P_{\rm j}  \over \pi r^2_{\rm diss} \Gamma^2 c } 
\left[ 1- {\Gamma\beta \over \Gamma_{\rm max}\beta_{\rm max} }
\left(1-\ \epsilon_{\rm B}\right)\right]
\end{equation}
The magnetic field $B$ scales as  $1/R_{\rm diss}$
both in the acceleration and in the coasting (i.e. $\Gamma=$const) 
phases, but with a different normalisation.


\subsection{Internal radiation}

This component corresponds mainly to the radiation produced by the blob
that can be efficiently scattered through the inverse Compton process.
This cannot be calculated without specifying the relativistic particle
distribution. 
If the emitting volume $V$ is a sphere, 
the synchrotron radiation energy density formally is:
\begin{eqnarray}
U^\prime_{\rm syn} \, &=& \, { V \over 4\pi r_{\rm diss}^2 c} m_{\rm e} c^2 
\int N(\gamma) \dot \gamma_{\rm syn} d\gamma
\nonumber \\
 &=& { 4 \over 9 } [n r_{\rm diss} \sigma_{\rm T}]  U^\prime_{\rm B} 
{ \int N(\gamma) \gamma^2 d\gamma \over \int N(\gamma) d\gamma }
\nonumber \\
 &=& { 4 \over 9 } [n r_{\rm diss} \sigma_{\rm T} \langle \gamma^2 \rangle ]  
U^\prime_{\rm B} 
\, = \, { 4 \over 9 }\, y \, U^\prime_{\rm B} 
\end{eqnarray}
where $\sigma_{\rm T}$ is the Thomson cross section, $n$ is the number 
density of the emitting particles and 
$y\equiv \sigma_{\rm T} n r_{\rm diss} \langle\gamma^2 \rangle$ is the relativistic 
Comptonization parameter.
This parametrisation follows the one in Ghisellini \& Tavecchio (2008).
From the fits to a sample of blazars performed in Celotti \& Ghisellini (2008)
we have that $y$ is in the range 0.1--10.
Since the found distribution of $y$ values is rather narrow, one
has a first order estimate of what is the internal radiation through the
value and profile of the magnetic energy density $U_{\rm B}$.

\begin{figure}
\hskip -0.2 cm
\psfig{figure=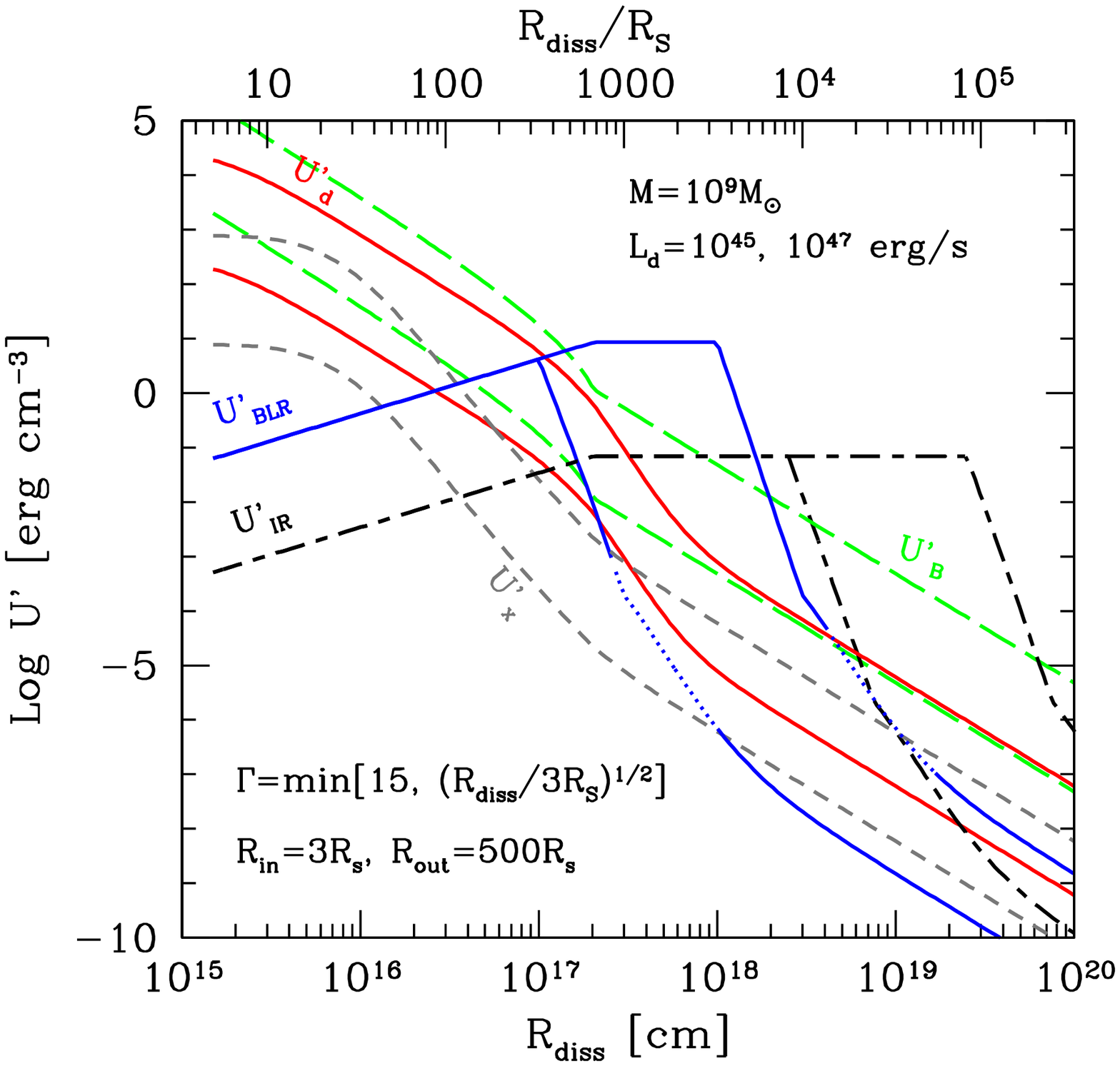,width=9cm,height=9.7cm}
\vskip -1 cm
\psfig{figure=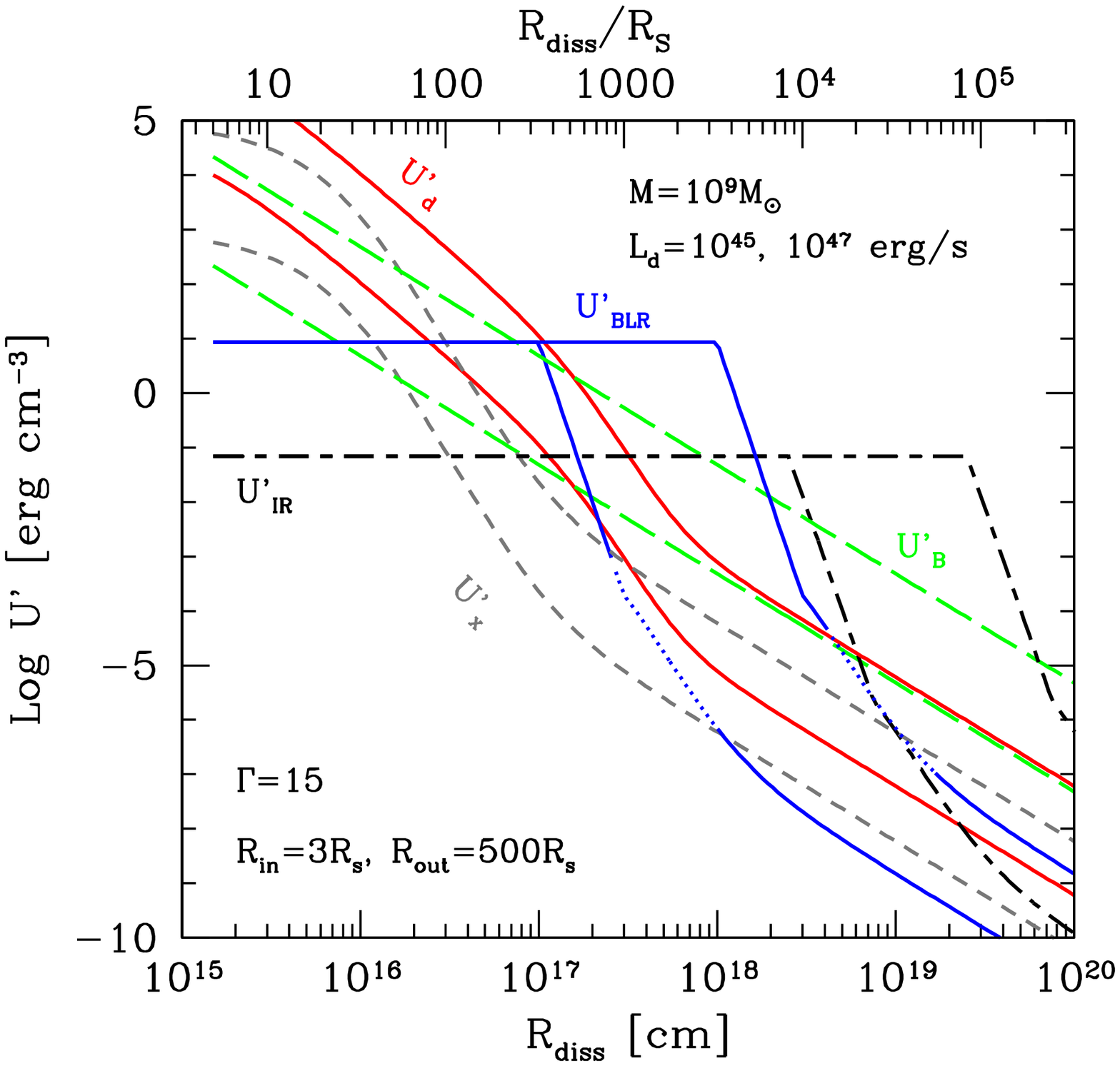,width=9cm,height=9.7cm}
\vskip -0.5 cm
\caption{
Top panel: Comparison of different energy densities 
as measured in the comoving frame.
The moving blob is assumed to have a bulk Lorentz factor
$\Gamma =\min[15, (R_{\rm diss}/3R_{\rm S})^{1/2}]$.
The black hole has a mass $M=10^9 M_\odot$. 
The different contributions are labelled.
The disk emits as a blackbody, and extends from 3 to 500 \sc\ radii.
The two sets of lines correspond to two disk luminosities:
$10^{45}$ and $10^{47}$ erg s$^{-1}$.
The radius of the broad line region is assumed to scale
with the disk luminosity as 
$R_{\rm BLR}=10^{17}L_{\rm d, 45}^{1/2}$ cm.
The radius of the IR torus scales as 
$R_{\rm IR} = 2.5\times 10^{18}L_{\rm d, 45}^{1/2}$ cm.
The X--ray corona is assumed to be homogeneous,
to extend up to 30 \sc\ radii and to emit 10 per cent of the 
disk luminosity.
The contribution of the BLR between 1 and 3 $R_{\rm BLR}$
depends on the unknown width of the BLR itself (dotted line). 
The magnetic energy density (long dashed lines) is calculated
assuming $P_{\rm j}=L_{\rm d}$ and $\epsilon_{\rm B}=0.1$.
Bottom panel: as above, but assuming a constant $\Gamma=15$
all along the jet.
}
\label{ur}
\end{figure}

\begin{figure}
\hskip -0.2 cm
\psfig{figure=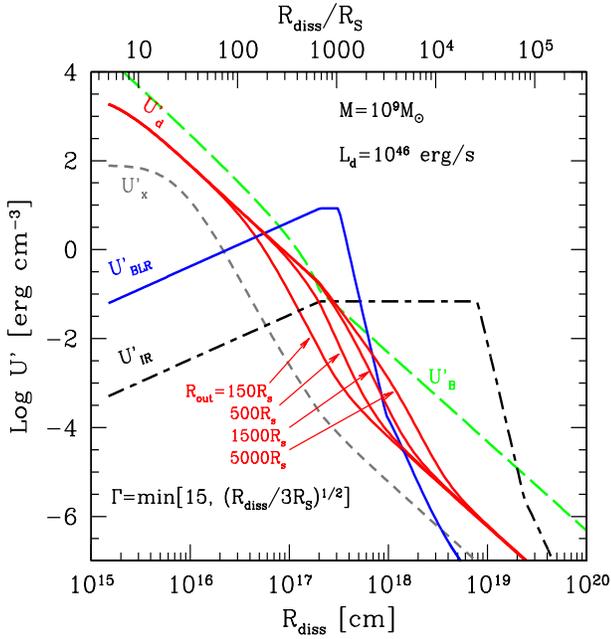,width=9cm,height=9.7cm}
\vskip -0.5 cm
\caption{
The energy density as seen in the comoving frame of the 
blob, moving with a bulk Lorentz factor
$\Gamma =\min[15, (R_{\rm diss}/3R_{\rm S})^{1/2}]$.
The black hole has a mass $M=10^9 M_\odot$, and the disk
emit $L_{\rm disk}=10^{46}$ erg s$^{-1}$.
The different contributions are labelled.
This figure shows the effect of changing the outer radius
of the accretion disk, from 150 to 5000 $R_{\rm S}$,
as indicated by the labels and arrows.
As the outer radius increases, $U^\prime_{\rm d}$ increases
at distances $R_{\rm diss}\sim R_{\rm out}$, but there the 
disk contribution is overtaken by $U^\prime_{\rm BLR}$
and $U^\prime_{\rm IR}$.
}
\label{urout}
\end{figure}

\subsection{Comparing the different components}

We have calculated the contributions to the energy density given by the 
different components as a function of $R_{\rm diss}$, 
plotting them in Fig. \ref{ur}.
For these cases, we have chosen a black hole
mass $M=10^9 M_\odot$, an accretion disk extending from 3$R_{\rm s}$ 
to 500$R_{\rm S}$, an accretion efficiency 
$\eta=0.08$, $f_{\rm BLR} = f_X=0.1$, $f_{\rm IR}=0.5$.
We plot the results for two values of $L_{\rm d}$: $10^{45}$ and $10^{47}$ 
erg s$^{-1}$. 
In the top panel we show the case of an accelerating jet, whose 
bulk Lorentz factor is $\Gamma=\min[15, (R_{\rm diss}/3R_{\rm S})^{1/2}]$.
In the bottom panel we assume a constant $\Gamma=15$ along the entire jet.
As can be seen, the dominant energy density is different at different 
$R_{\rm diss}$.
As a rule, all the external radiation energy densities drop when 
$R_{\rm diss}$ is greater than the corresponding typical size of 
the structure producing the seed photons.

Particularly interesting is the comparison between 
$U^\prime_{\rm d}$ and $U^\prime_{\rm BLR}$.
The distance above which $U^\prime_{\rm BLR}>U^\prime_{\rm d}$
depends upon the disk luminosity, since the BLR adjusts its radius so 
to give a constant (in the lab frame) energy density.
So for $L_{\rm d}=10^{47}$ erg s$^{-1}$ the energy density from the BLR dominates
only above $10^{17}$ cm, equivalent to $\sim 300 R_{\rm S}$, 
while for $L_{\rm d}=10^{45}$ erg s$^{-1}$ it starts to dominate about 
three times closer, when $\Gamma$ has not reached yet its maximum value.
Note that, for the shown cases, $U^\prime_{\rm B}$ dominates
over the external radiation energy density only
at the start of the jet and up to 100--300 $R_{\rm S}$,
where $U^\prime_{\rm BLR}$ takes over.
In this particular examples, we assumed $P_{\rm B}=0.1 L_{\rm d}$.

Fig. \ref{urout} illustrates the effects of changing the outer radius of 
the accretion disk, from $R_{\rm out}=150$ to $5000$ \sc\ radii.
As expected, $U^\prime_{\rm d}$ remains the same at small and very high
distances, and increases with $R_{\rm out}$ in between, at distances
comparable with $R_{\rm out}$.
The increase of $U^\prime_{\rm d}$ is modest, and occurs when the 
external radiation energy density is dominated by the BLR and torus
components.

\begin{figure}
\vskip -0.5cm
\hskip -0.2cm
\psfig{figure=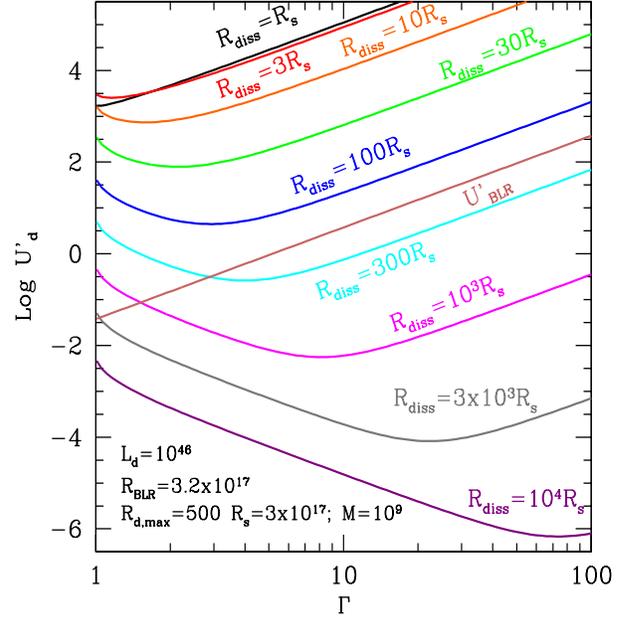,width=9cm,height=10cm}
\vskip -1 cm
\caption{
The radiation energy density $U^\prime_{\rm d}$
(as seen in the comoving frame) produced by the direct radiation 
of the accretion disk as a function of $\Gamma$, for different 
distances $R_{\rm diss}$, for a total disk luminosity $L_{\rm d}=10^{46}$ 
erg s$^{-1}$ and a black hole mass $M=10^9 M_\odot$.
For comparison we show the radiation energy density observed {\it within} the BLR.
It can be directly compared with $U^\prime_{\rm d}$ 
only for $R_{\rm diss}<R_{\rm BLR}$. 
}
\label{ug}
\end{figure}

In Fig. \ref{ug} we show $U^\prime_{\rm d}$ as a function 
of $\Gamma$ for different locations of the dissipation region.
Of course the largest $U^\prime_{\rm d}$ are obtained when
the source is very close to the accretion disk.
For each $R_{\rm diss}$ there is a value of $\Gamma$ 
that minimises $U^\prime_{\rm d}$.
To understand this (somewhat anti--intuitive) behaviour
we must recall that the boosting factor $b=\Gamma(1-\beta\mu)$
has a minimum for $\beta=\mu$.
This corresponds, in a comoving frame, to $\mu^\prime=0$, i.e. an angle 
of $90^\circ$.
Therefore, when the external radiation is all coming from a ring,
we expect that in the comoving frame, the radiation energy density
initially decreases increasing $\beta$, but when the photons
are starting to come from the forward hemisphere (i.e. $\mu^\prime<0$,
or, equivalently, $\mu<\beta$), $U^\prime_{\rm d}$ increases increasing $\beta$.
This explains the minima shown in Fig. \ref{ug}, once we 
weight the above effect with the intensity produced at each ring.

\begin{figure}
\vskip -0.5cm
\hskip -0.2cm
\psfig{figure=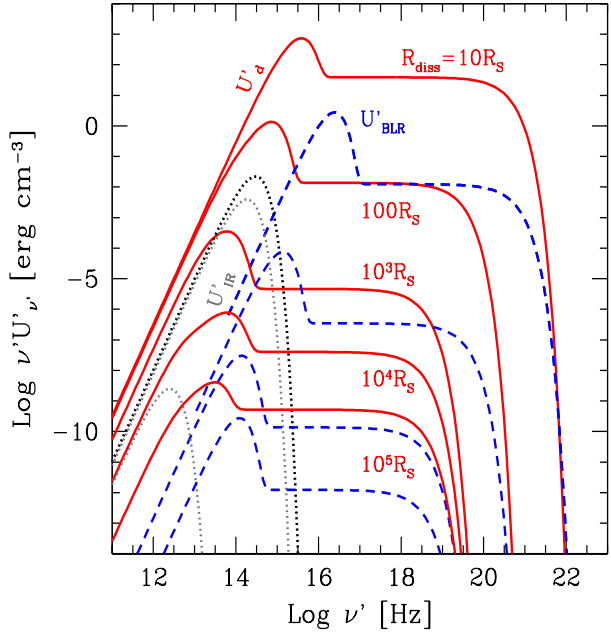,width=9cm,height=10cm}
\vskip -1 cm
\caption{
The spectra of the radiation energy density (as seen in the 
comoving frame) produced by the direct disk plus the
X--ray corona  ($U^\prime_{\rm d}$, solid line),
the BLR ($U^\prime_{\rm BLR}$, including the ``reflected'' X--rays, 
dashed line) and the IR torus ($U^\prime_{\rm IR}$, dotted line).
We have assumed a constant $\Gamma=15$, $M=10^9 M_\odot$ and 
$L_{\rm d}=10^{45}$ erg s$^{-1}$.
The spectra are shown for 5 different $R_{\rm diss}$.
For $R_{\rm diss}=10$ and 100 \sc\ radii  $U^\prime_{\rm BLR}(\nu^\prime)$ 
is the same, since for these cases $R_{\rm diss}<R_{\rm BLR}$.
The contribution of the IR emitting torus is the same up
to  $R_{\rm diss}=10^3 R_{\rm S}$, and starts to decline
afterwards, when $R_{\rm diss}>R_{\rm IR}$.
}
\label{uradz}
\end{figure}

\subsection{Spectra of the external radiation}

Besides comparing the frequency integrated components
forming the radiation energy density in the comoving frame, it is
instructive to compare also their different spectra.
This is because the Doppler boost appropriate for each
component is not the same, and it changes for different 
distances from the black hole.
In Fig. \ref{uradz} we show one illustrative example, calculated
for a blob moving with a constant $\Gamma=15$ and located at different $R_{\rm diss}$.
The disk is assumed to emit $L_{\rm d}=10^{45}$ erg s$^{-1}$ and the black hole
mass is $M=10^9 M_\odot$, corresponding to $R_{\rm S}=3\times 10^{14}$ cm.
With this disk luminosity, the BLR is located at $R_{\rm BLR}=10^{17}$ cm,
implying that for $R_{\rm diss}< R_{\rm BLR}$ the radiation energy density
from the BLR is the same.
For larger $R_{\rm diss}$ the total $U^\prime_{\rm BLR}$ decreases 
(see Eq. \ref{ublr2}).
The radiation energy density from the disk and its corona 
always decreases increasing $R_{\rm diss}$, being larger than
$U^\prime_{\rm BLR}$ for  $R_{\rm diss}< 100R_{\rm S}\sim 3\times 10^{16}$ cm
(see also Fig. \ref{ur}). 
The disk radiation dominates again for $R_{\rm diss}\gsim \times 10^{18}$ cm
(see Fig. \ref{ur}).
For small $R_{\rm diss}$, the high energy X--ray photons coming
from the corona will be responsible of a large optical depth 
for the $\gamma$--$\gamma$ process, as discussed below.
The IR energy density coming form the torus is constant
for all $R_{\rm diss}$ but for the largest value
(since $R_{\rm IR}=8.3\times 10^3 
R_{\rm S}=2.5\times 10^{18}$ cm).
This component will be larger than $U^\prime_{\rm d}$ and
$U^\prime_{\rm BLR}$ for $R_{\rm diss}$ greater than $\sim 500R_{\rm S}
\sim 2\times 10^{17}$ cm.

\subsection{Input parameters}

Having included many components of external radiation,
and having tried to link the jet emission to the
accretion disk luminosity and the black hole mass,
we necessarily have a large numbers of parameters
for a specific model.
It can be useful to summarise them here, dividing them into
two separated lists.
In the latter we have those parameters that we fix for
all models, based on physical considerations or 
a priori knowledge, or that do not influence the model SED
(apart from pathological cases); in the former we list
the important input parameters.
\begin{itemize}
\item 
$M$: the mass of the black hole;

\item 
$L_{\rm d}$: the luminosity of the accretion disk.

\item 
$R_{\rm diss}$: the distance from the black hole where  the
jet dissipates.

\item 
$P^\prime_{\rm i}$: the power, calculated in the jet rest frame,
injected into the source in the form of relativistic electrons.

\item 
$B$: the magnetic field on the dissipation region;

\item
$\Gamma_{\rm max}$: the value of bulk Lorentz factor after the acceleration
phase, in which the bulk Lorentz factor increases as 
$[R/(3R_{\rm S})]^{1/2}$.

\item
$\theta_{\rm v}$: the viewing angle.

\item 
$\gamma_{\rm b}$: value of the random Lorentz factor at which
the injected particle distribution $Q(\gamma)$ changes slope.

\item 
$\gamma_{\rm max}$: value of maximum random Lorentz factor of 
the injected electrons.

\item 
$s_2$: slope of $Q(\gamma)$ above $\gamma_{\rm b}$.

\end{itemize}

These amount to 10 parameters. 
If we use $\theta_{\rm v}\sim 1/\Gamma$, as discussed in \S \ref{caveats}, 
then the number of relevant parameters decreases to 9.
Note that we do not consider $P_{\rm j}$ as an input parameter,
since it is derived once calculating the total electron energy density
in the dissipation region, and assuming a given number of (cold) proton
per emitting electron.



The following parameters are either unimportant or are well constrained:

\begin{itemize}

\item 
$R_X$: the extension of the X--ray corona. 
We fix it to $30R_{\rm S}$.

\item $R_{\rm out}$: the outer radius of the accretion disk. 
For the shown illustrative examples, we fixed it at 500 \sc\ radii,
and we have shown that changing it has a modest influence on $U^\prime_{\rm d}$.

\item 
$s_1$: slope of $Q(\gamma)$ below $\gamma_{\rm b}$.
It is bound to be very flat (i.e. $-1<s_1<1$).

\item
$\alpha_X$: the spectral shape of the coronal X--ray flux.
We fix it to $\alpha_X=1$.

\item 
$h\nu_{\rm c}$: the high energy cut--off of the X--ray coronal flux.
We fix it at 150 keV.

\item
$L_X$: total X--ray luminosity of the corona.
We fix it to $L_X=0.3L_{\rm d}$.

\item
$f_{\rm BLR}$: fraction of $L_{\rm d}$ intercepted by the BLR and
re--emitted in broad lines. 
We fix it to $f_{\rm BLR}=0.1$

\item
$f_{\rm BLR,X}$: fraction of $L_X$ of the corona scattered by the BLR.
We fix it to $f_{\rm BLR,X}=0.01$

\item
$f_{\rm IR}$: fraction of $L_{\rm d}$ intercepted by the torus
and re--emitted in IR. 
We fix it to $f_{\rm IR}=0.5$.

\item
$\psi$: semi--aperture angle of the jet.
We fix it to $\psi=0.1$.

\item $U_{\rm star}$: this is the radiation energy density within the bulge of the
host galaxy. We approximated it with the constant value  $U_{\rm star}=10^{-10}$ erg cm$^{-3}.$

\end{itemize}

\section{The role of pairs}

At large distances from the black hole, the density
of the photon targets for the $\gamma$--$\gamma \to e^\pm$ process
is small, with few pairs being produced inside the emitting region.
At small distances, instead, the presence of the X--ray corona 
makes pairs easy to create.
With our assumptions, we can see the effects of the reprocessing of the
spectrum due to pairs.
Whenever pairs are produced
inside the emitting region at small $R_{\rm diss}$, 
the radiative cooling rates are large,
implying that all particles cool in one light crossing time.
This in turn implies a particle distribution of the form given 
by Eq. \ref{ngamma}, with $\gamma_{\rm cool}\sim 1$.

One necessary ingredient for making pairs important is the relative
amount of power emitted above the energy threshold $m_ec^2$ (in the comoving frame).
If this power is a small fraction of the total, the
reprocessing will be modest, even if all the $\gamma$--rays get absorbed.
There is then the possibility to have a large jet power dissipated close 
to the accretion disk, but with electrons of lower energies, whose emission
is mostly below threshold.
Furthermore, even if the electron energies are large, the bulk Lorentz
factor could be small, especially very close to the accretion disk,
where presumably the jet just starts to accelerate.
In this case the magnetic energy density becomes more dominant 
with respect to the disk radiation, implying a relatively modest high energy
emission, and furthermore the whole spectrum is much less boosted.

In the following we will consider the case of two jets
with the same parameters, except that the first is moving with
a large $\Gamma$ already at small $R_{\rm diss}$, 
the other, instead, is accelerating.

\begin{figure}
\vskip -0.5cm
\psfig{figure=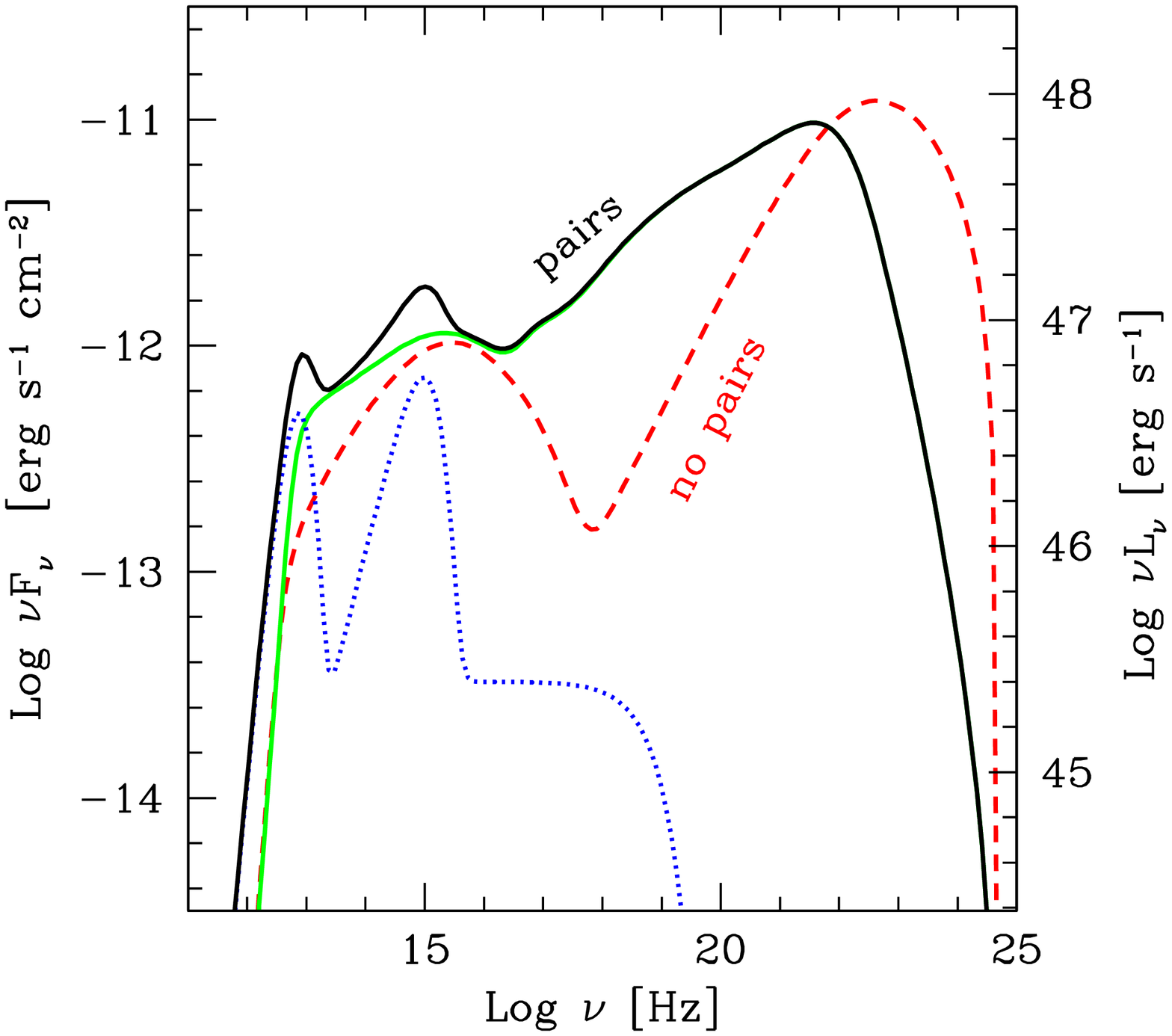,width=9cm,height=8cm}
\vskip -0.5cm
\psfig{figure=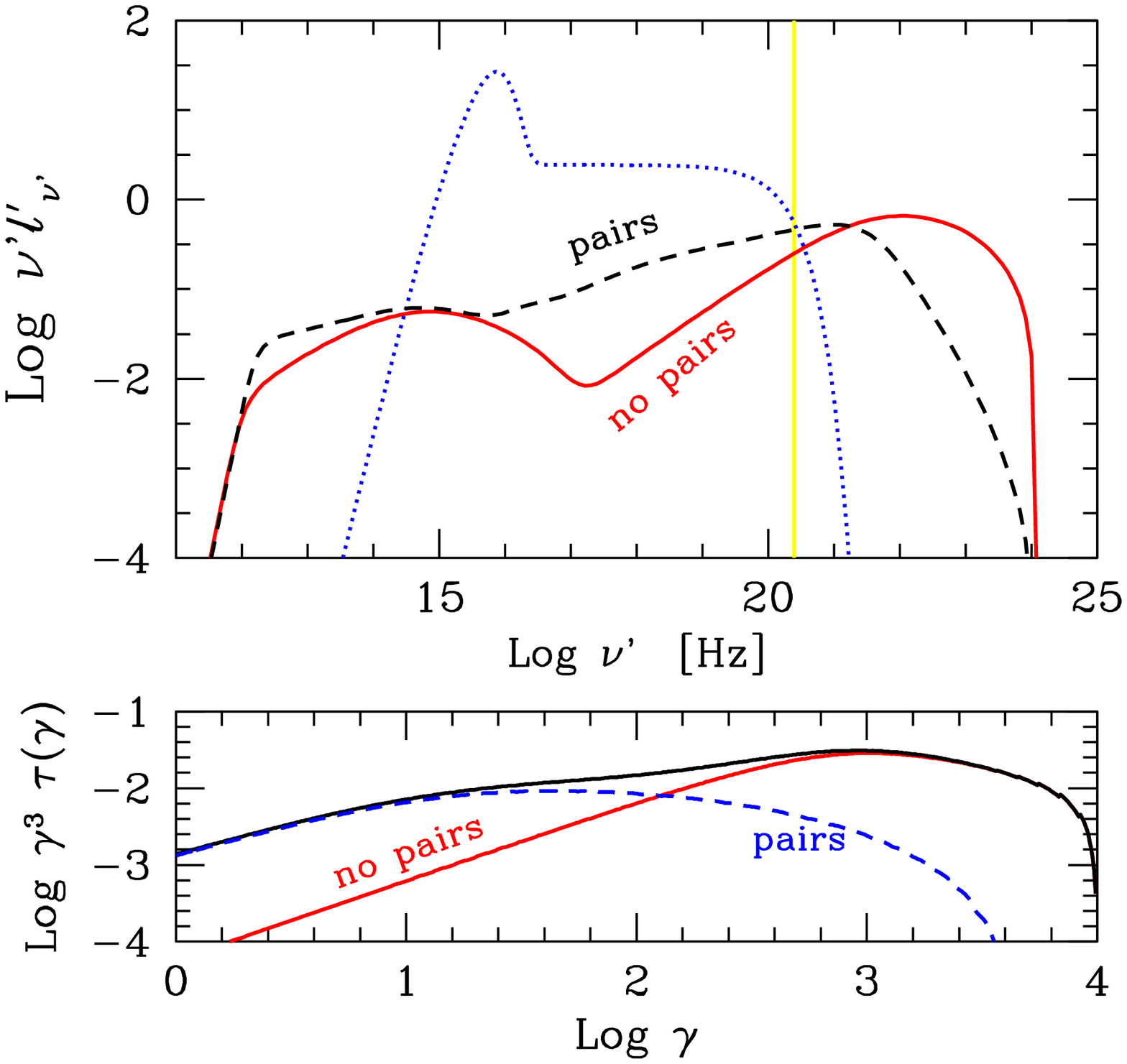,width=9.cm,height=10.5cm}
\vskip -0.5cm
\caption{
Top panel:
example of how reprocessing due to electron--positron pairs
can be important, if the dissipation region is located too close 
to the accretion disk and its X--ray corona, {\it and} if this
region produces $\gamma$--rays above the pair--production threshold.
The two humps at low energies correspond to the flux produced by the
accretion disk and the IR flux from the torus.
See Tab. \ref{para} for the parameters used to construct the shown SED.
Mid panel: the SED as observed in the comoving frame.
Bottom panel: 
the particle distribution $\gamma^3\tau(\gamma)$.
For this example, the maximum particle Lorentz factor is 
$\gamma_{\rm max}=10^4$ and $\gamma_{\rm b}=10^3$.
A redshift $z=3$ has been assumed.
}
\label{pairs}
\end{figure}

Consider Fig. \ref{pairs}: it
shows the effects of including the $\gamma$--$\gamma \to e^\pm$
on the final spectrum.
In the middle panel of  Fig. \ref{pairs} the same spectra are shown as seen in 
the comoving frame.
In this case on the y--axis we plot the specific compactness, defined as 
$\ell^\prime(\nu^\prime) \equiv 
\sigma_{\rm T} L^\prime(\nu^\prime)/ (r_{\rm diss} m_e c^3)$.
The bottom panel of this figure shows the particle distribution
(considering or not the produced pairs) in the form $\gamma^3\tau(\gamma)\equiv
\gamma^3 \sigma_{\rm T} r_{\rm diss} N(\gamma)$.
The $\gamma^3$ factor makes the $\gamma^3\tau(\gamma)$ distribution to have
a peak: electrons at this peak are the ones producing the peaks in the 
$\nu L_\nu$ synchrotron and inverse Compton spectra.
The parameters used for this example are listed in Tab. \ref{para}.

One can see that pairs redistribute the power from high to low frequencies.
The $N(\gamma)$ distribution steepens at low energies (where
pairs contribute the most), and this softens the emitted spectrum
especially in the X--ray range.
The reprocessing of the X--ray spectrum
is stronger than at lower frequencies because the low energy
electrons emit, by synchrotron, self absorbed radiation, whose
shape and normalisation are largely independent of the slope of 
$N(\gamma)$.
Instead, in the X--ray range, the presence of the intense optical--UV
radiation field coming from the accretion disk implies that the dominant 
process is the external Compton (EC) scattering with these photons.
In turn, this implies that the radiation we see in the soft and hard
X--ray range is due to relatively low energy electrons.
In fact, the scattered frequency $\nu \sim \gamma^2\Gamma^2\nu_0$,
where $\nu_0$ is the (observed) frequency of the seed photon.
Therefore, at 10 keV, we are observing the radiation produced by electrons
with $\gamma \sim 50 \, \nu_{0,15}^{-1/2}\Gamma^{-1} \sim$ a few.

\begin{figure}
\hskip -0.2 cm
\vskip -0.3 cm
\psfig{figure=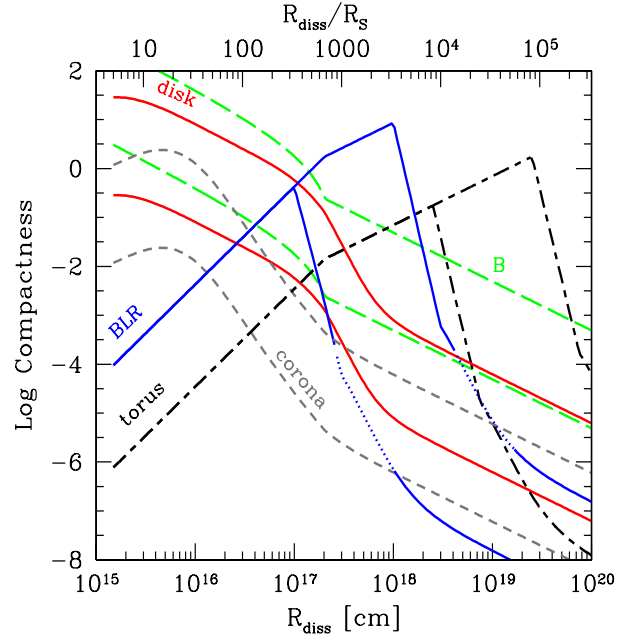,width=9cm,height=9.7cm}
\vskip -0.5cm
\caption{
The compactness 
$\ell^\prime \equiv 4\pi U^\prime \sigma_{\rm T} r_{\rm diss}/ (m_{\rm e}c^2)$  
as measured in the comoving frame of a source of size 
$r_{\rm diss}$ as a function of $R_{\rm diss}$.
We use the same parameters used to construct the top panel 
of Fig. \ref{ur}.
}
\label{elle}
\end{figure}

In the above example we have, on purpose, assumed that
the dissipation region, although very close to the
black hole, is already moving at large speeds,
corresponding to a bulk Lorentz factor of 10.
We may ask, instead, what happens if the same kind of 
dissipation occurs while the jet is still accelerating,
and has therefore a relatively small bulk velocity when it is
close to the black hole.

To this aim, Fig. \ref{elle} shows the compactness $\ell^\prime$, 
as measured in the comoving frame, as a function of $R_{\rm diss}$,
for an accelerating jet.
We define this comoving compactness as:
\begin{equation}
\ell^\prime\,  \equiv \, {4\pi U^\prime \sigma_{\rm T} r_{\rm diss}
\over m_{\rm e}c^2 }
\end{equation}
where $r_{\rm diss}$ is the size of the blob.
The term $U^\prime$ is the external radiation energy densities.
The compactness is directly associated to the optical depth of the 
$\gamma$--$\gamma \to e^\pm$ process, and it describes the absorption
probability of a $\gamma$--ray photon while it is inside the blob.
In other words, this definition does not account for the absorption
a photon can suffer while it has already escaped the blob.
It also neglects the contribution of the internal radiation energy
density due to the synchrotron and the synchrotron self Compton (SSC)
emission.
Fig. \ref{elle} shows that there are three distances where the
compactness is large:
\begin{enumerate}
\item 
At the base of the jet the main targets
for the $\gamma$--$\gamma$ process are the X--ray photons
produced by the corona. 
They can interact with $\gamma$--ray photons just above threshold,
implying that the absorption is important. However, at these distances,
the magnetic energy density is larger than the external radiation energy 
density (dominated by the disk radiation), and this implies a modest 
external Compton scattering, and a modest production of $\gamma$--rays.

\item
The compactness becomes large again for $R_{\rm diss}\sim R_{\rm BLR}$.
Since the main contribution to $U^\prime$ is due to the BLR photons,
(seen at UV frequencies in the comoving frame), the absorbed $\gamma$--rays
will have GeV energies (see next subsection for more details).

\item
The third relevant range of distances corresponds to 
$R_{\rm diss}\sim R_{\rm IR}$. 
The main targets for the absorption process are the IR photons produces 
by the torus, absorbing TeV $\gamma$--rays.
\end{enumerate}

\begin{figure}
\vskip -0.5cm
\psfig{figure=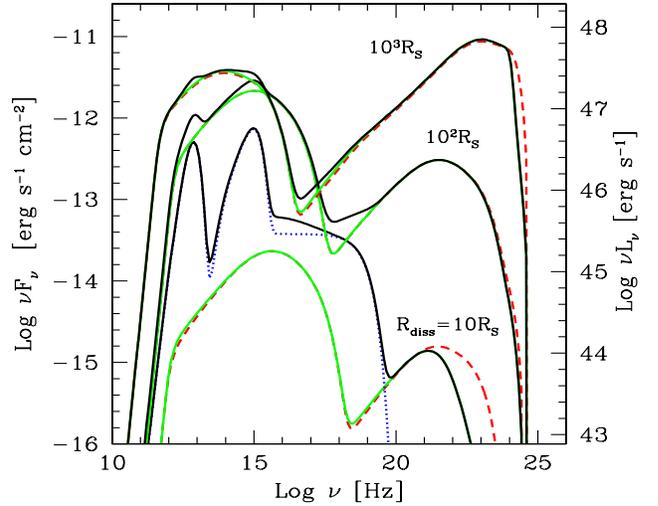,width=9cm,height=8cm}
\vskip -0.5cm
\caption{
Same as Fig. \ref{pairs}, but for an accelerating jet and 
for different $R_{\rm diss}$.
In this case, in the comoving frame of the jet, the magnetic energy 
density is dominant for $R_{\rm diss}=$10 and 100$R_{\rm S}$.
Furthermore, the much reduced Doppler boosting for small values
of $R_{\rm diss}$ implies a very small received flux.
There is some $\gamma$--$\gamma$ absorption (compare solid and 
dashed lines, the latter corresponding to switching off the 
$\gamma$--$\gamma$ process), but involving a very small amount of power.
Consequently, pairs reprocessing is unimportant.
Light grey lines (green in the electronic version) correspond to the 
non--thermal spectrum, black lines include the contribution from the 
accretion disk, its X--ray corona and the emission from the IR torus.
}
\label{rpairs}
\end{figure}

Fig. \ref{rpairs} shows three SED corresponding to 
$R_{\rm diss}=10$, 100 and $10^3R_{\rm S}$, with an
electron injection function equal to the one used in Fig. \ref{pairs},
but assuming $\Gamma=\min[15, (R_{\rm diss}/3R_{\rm S})^{1/2}]$.
Tab. \ref{para} lists the parameters.
We can see the dramatic changes with respect to Fig. \ref{pairs},
and also among the three shown models.
Pair production, although present (compare solid and dashed lines),
is marginal and there is no noticeable reprocessing of the primary spectrum.
This is due to the much reduced importance of the external radiation
when $\Gamma$ is small. 
As a consequence, the spectrum is dominated by the synchrotron emission,
with a small fraction of the power being emitted above
the pair production energy threshold.
This in turn implies that early (i.e. close to the black hole)
dissipation is not an efficient mechanism to produce 
electron--positron pairs.
Since a small $\Gamma$ means small Doppler boosting, the
produced spectra at small distances are undetectable, 
overwhelmed by the emission of more distant and fastly
moving components.

When $\Gamma$ is large, therefore for the $R_{\rm diss}=10^3R_{\rm S}$ case,
the relevant absorbing photons are those of the broad line region, 
with a very modest contribution from the coronal X--rays reflected by the BLR
material. 
The overall effect is marginal.

We can conclude that a ``canonical" jet can dissipate 
part of its kinetic energy even at small distances.
One way to avoid strong pair production is to have
small electron energies, emitting a small power above
the pair production threshold.
Alternatively, if the jet is accelerating, and close to
the black hole it has a modest $\Gamma$--factor, 
the produced radiation is hardly observable because
the Doppler boost is limited.
In relative terms, the synchrotron luminosity should be
dominant, with a small fraction of power being emitted
at high energies.

\subsection{Pair production versus Klein--Nishina effects}

In powerful blazars, the high energy flux, at $\sim$GeV energies, is
dominated by the inverse Compton process between high energy electrons
and external radiation.
In Tavecchio \& Ghisellini (2008) we pointed out that when the dominant
contribution to the seed photon for scattering is given by the BLR, the
hydrogen Lyman--$\alpha$ photons are the most prominent ones.
The fact that there is a characteristic frequency of the seed photons
allows to easily calculate when the Klein--Nishina effects 
(decrease of the scattering cross section with energy) are important.
For completeness, we briefly repeat here the argument, with the
aim to extend it to cases in which $R_{\rm diss}$ is beyond the BLR, and
as a consequence, the relevant seed photons becomes the IR ones.

In the comoving frame, and as long as we are within the BLR, 
The observed seed Lyman--$\alpha$ frequency is observed at
$\nu_{\rm L\alpha}^\prime=2\Gamma \nu_{\rm L\alpha}$. 
To be in the Thomson scattering regime we require that
\begin{equation}
2\Gamma h\nu_{\rm L\alpha}\, < \, { m_{\rm e}c^2\over \gamma}
\end{equation}
If the random Lorentz factor $\gamma$ of the electron satisfies this 
condition, the energy of the scattered photon is
\begin{equation}
\nu_{\rm KN} \, =\, {4\over 3}\gamma^2  \nu_{\rm L\alpha}
{2 \Gamma \delta \over 1+z} \, =\, 
15\, {\delta \over \Gamma (1+z)}  \,\, {\rm GeV}
\label{kn}
\end{equation}
Above this energy, the spectrum steepens due to the 
decreased efficiency of the scattering process. 
This may well mimic the effect of photon--photon absorption,
making it difficult to discriminate between the two effects.
Consider also that the Lyman--$\alpha$ photons are the
best targets for the photon--photon process at these energies
and in these conditions (namely, relatively far from the accretion disk, 
but still within the BLR).

We can repeat the very same argument assuming that the relevant seed
photons are the ones produced by the IR torus. This occurs when the 
dissipation region is beyond the BLR but within $R_{\rm IR}$.
In this case the relevant frequency is $\nu_{\rm IR}=3\times 10^{13}$ Hz
and we can rewrite Eq. \ref{kn} as
\begin{equation}
\nu_{\rm KN} \, =\, {4\over 3}\gamma^2  \nu_{\rm IR}{2 \Gamma \delta \over 1+z} \, =\, 
1.2\, {\delta \over \Gamma (1+z)}  \,\, {\rm TeV}
\label{kn2}
\end{equation}
This implies that, if we do see a spectrum which is unbroken even above $15/(1+z)$ GeV,
we can conclude that the dissipation region is beyond the BLR.

Also in this case, the ``intrinsic" absorption due to the $\gamma$--$\gamma$ process
is expected to become important at $\sim \nu_{\rm KN}$.
There is therefore a range, between $15/(1+z)$ and $\sim 1000/(1+z)$ GeV,
in which the intrinsic spectrum should not suffer from the Klein--Nishina
nor from the $\gamma$--$\gamma$ process due to ``internal" absorption.
These sources are thus the best candidates for studying the 
photon--photon absorption due to the cosmic optical--UV background radiation.

\section{Dissipation at large distances}

Although there are strong indications that {\it most} of the 
observed radiation originates in a well localised region of the
jet, there is also compelling evidence that the jet dissipates
also in other regions, larger and more distant from its base.
In fact, in the radio band, we do see bright knots at different
distances contributing to form the characteristic flat total 
radio spectrum of the entire jet, and we do observed optical 
and X--ray knots also at hundreds of kpc from the jet apex.
The power associated to these, often resolved, features is
a small fraction of the total observed jet bolometric luminosity,
but in specific bands the flux produced at larger distances can
be comparable to the flux produced in the most active region.

It is therefore instructive to calculate the predicted spectrum
as a function of distance, extending our analysis well beyond
the region of influence of the IR flux produced by the torus,
and therefore analysing the effects of the radiation energy 
density of the cosmic background radiation (CMB). 
This has already been pointed out as an important source of
seed photons for enhancing the inverse Compton flux of the 
bright X--ray knots detected by the Chandra satellite at tens and
hundreds of kpc from the core of the jet (Tavecchio et al. 2000;
Celotti Ghisellini \& Chiaberge 2001).

\begin{figure}
\vskip -0.7cm
\hskip -1.6cm
\psfig{figure=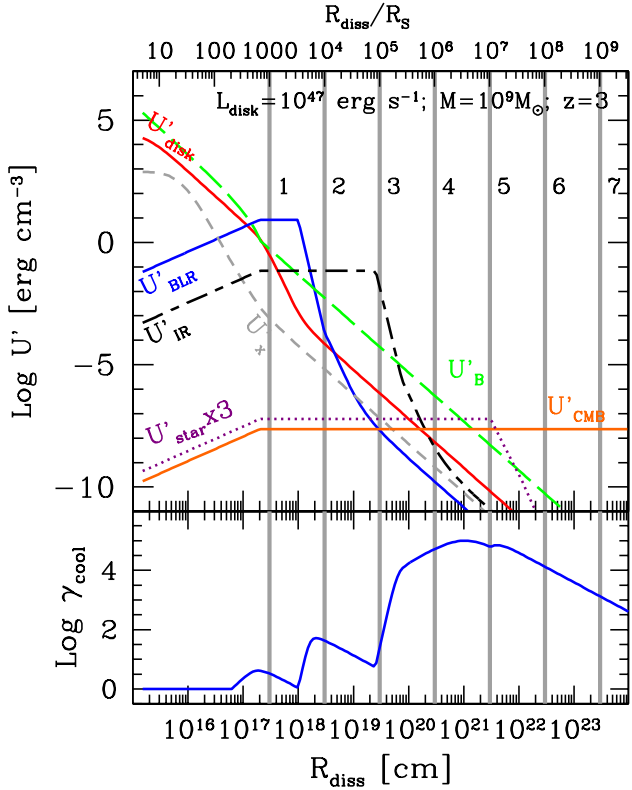,width=12cm,height=12cm}
\vskip -0.5 cm
\caption{
Top panel: the contributions to the radiation energy density as 
seen in the comoving frame of the emitting blob, as labelled, and the
magnetic energy density as a function of the distance of the blob
to the black hole.
We assumed: $M=10^9 M_\odot$, 
$L_{\rm d}=10^{47}$ erg s$^{-1}$, $z=3$ 
and $\Gamma =\min[15, (R_{\rm diss}/3R_{\rm S})^{1/2}]$.
The Poynting flux is $P_{\rm B}=10^{46}$ erg s$^{-1}$ once the jet
has reached its maximum bulk Lorentz factor.
The grey vertical lines indicates the distances assumed to construct
the SEDs in Fig. \ref{rdiss}, the different numbers help to identify
the corresponding SED.
The bottom bottom panel shows $\gamma_{\rm cool}$ (after one light crossing 
time since the start of the injection) as a function of $R_{\rm diss}$. 
}
\label{cmb}
\end{figure}

Fig. \ref{cmb} shows the different contributions to the radiation energy 
densities as seen in the comoving frame, compared to $U^\prime_{\rm B}$,
and including $U^\prime_{\rm CMB}$, important for large distances.
To construct this figure we have assumed $M=10^9 M_\odot$, 
$L_{\rm d}=10^{47}$erg s$^{-1}$, 
$P^\prime_{\rm i}=10^{44}$ erg s$^{-1}$, $P_{\rm B}=10^{46}$ erg s$^{-1}$, 
$z=3$ and
$\Gamma =\min[15, (R_{\rm diss}/3R_{\rm S})^{1/2}]$.
This figures shows that $U^\prime_{\rm B}$ dominates only in two
regions of the jet: in the vicinity of the black hole and 
at $R_{\rm diss}\sim 10^{20}$--$10^{21}$ cm.
Within these two distances, $U^\prime_{\rm BLR}$ and 
$U^\prime_{\rm IR}$ dominate, and beyond $10^{21}$ cm
$U^\prime_{\rm CMB}$ takes over.
%
%
This has two important and immediate consequences:
\begin{enumerate}
\item The external Compton radiation will be more important than
the synchrotron (and the SSC, if $y\sim 1$) 
radiation at all distances, except the (relatively narrow) 
distance intervals where $U^\prime_{\rm B}$ dominates.

\item 
At large distances the jet is conical,
and the blob size scales linearly with $R_{\rm diss}$
(namely the jet is a cone with the same aperture angle $\psi$). 
Then the light crossing time $t_{\rm cross}=\psi R_{\rm diss}/c$.
This hypothesis means that $\gamma_{\rm cool}$ reaches a 
maximum where $U^\prime_{\rm B}=U^\prime_{\rm CMB}$.
\end{enumerate}
To understand the second issue, consider 
the value of $\gamma_{\rm cool}$ after a time $t_{\rm cross}$.
\begin{equation}
\gamma_{\rm cool} \, = 
\, {3 m_{\rm e} c^2 \over 4\sigma_{\rm T} r_{\rm diss} U^\prime}  
\label{gcool1}
\end{equation}
In the regions of interest (i.e. above 1 kpc),
$U^\prime_{\rm CMB}$ is constant, while 
$U^\prime_{\rm B}\propto r^{-2}_{\rm diss}$.
Therefore the maximum $\gamma_{\rm cool} $ occur when 
$U^\prime_{\rm CMB}=U^\prime_{\rm B}$, i.e. at
\begin{equation}
\psi R_{\rm eq} \, = \, \left[ { P_{\rm B} \over 
\pi a c}\right]^{1/2} \, { 1\over T_0^2 (1+z)^2 \Gamma^2} 
\end{equation}
Inserting this in Eq. \ref{gcool1} and setting 
$U^\prime=U^\prime_{\rm CMB}+U^\prime_{\rm B}
=2 U^\prime_{\rm CMB}$, we obtain:
\begin{eqnarray}
\gamma^{\rm max}_{\rm cool} \, &=& \, 
{3 m_{\rm e} c^2 \over 4\sigma_{\rm T} T_0^2(1+z)^2} 
\,\left( { \pi c \over a  P_{\rm B}}\right)^{1/2}  \nonumber \\
& =& \, { 3.08\times 10^6 \over (1+z)^2 \,  P_{\rm B,46}^{1/2}}
\label{gcool2}
\end{eqnarray}
Independent of $\Gamma$. 
Electrons with this energy emit an observed synchrotron frequency 
$\nu_{\rm cool}$ given by:
\begin{eqnarray}
\nu^{\rm syn}_{\rm cool} \, &\sim& \, 3.6 \times 10^6 
B\gamma_{\rm cool}^2 \delta  \nonumber \\
&=&\,  1.09 \times 10^{14}\, { \Gamma\delta \over 
(1+z)^3 L_{\rm B,46} }\quad {\rm Hz}
\label{vcoolsyn}
\end{eqnarray}
Analogously, the observed frequency emitted by these electrons scattering the peak
of the CMB radiation is
\begin{eqnarray}
\nu^{\rm IC}_{\rm cool} \, &\sim& \, {4 \over 3} {3.93 kT_0 \over h} \gamma_{\rm cool}^2 
\Gamma\delta  \nonumber \\
&=&\,  2.8 \times 10^{24}\, { \Gamma\delta \over 
(1+z)^4 P_{\rm B,46} }\quad {\rm Hz}
\label{vcoolic}
\end{eqnarray}
We thus expect that the synchrotron spectrum produced in 
powerful jets, at large distances, cuts--off at a frequency 
given by Eq. \ref{vcoolsyn}, or somewhat larger if, at these
scales, the active region has a size smaller than $\psi R$
(implying a smaller cooling time, and thus electron energies
greater than the ones given by Eq. \ref{gcool2}).

\begin{figure}
\vskip -0.5cm
\psfig{figure=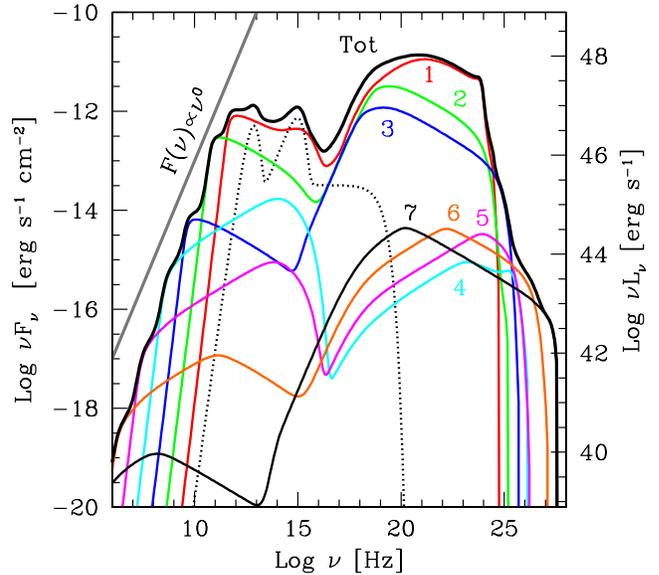,width=9cm,height=9cm}
\vskip -0.5cm
\caption{
Sequence of SEDs calculated for different $R_{\rm diss}$ 
from $10^3 R_{\rm S}$ to $10^9 R_{\rm S}$ (one per decade).
The injected electron luminosity is $P^\prime_{\rm i}=10^{44}$ erg s$^{-1}$
for $R_{\rm diss}=10^3 R_{\rm S}$ and is reduced by a factor 3 each decade.
The particle distribution has always the same $\gamma_{\rm b}=100$, while 
$\gamma_{\rm max}$ increases by a factor 3 each decade starting
from $\gamma_{\rm max}=10^4 $ for $R_{\rm diss}=10^3 R_S$.
All other parameters are the same as in Fig. \ref{cmb}.
What shown are the observed spectra neglecting the absorption of the 
high energy flux due to the IR--opt--UV cosmic background.
The numbers correspond to the same numbers in Fig. \ref{cmb} and 
correspond to the SED at different $R_{\rm diss}$.
The thicker black line is the sum of all the SED.
The grey line at radio frequencies indicates $F(\nu) \propto \nu^0$.
The received flux is calculated assuming that the source is at $z=3$.
}
\label{rdiss}
\end{figure}

Fig. \ref{rdiss} shows the predicted SEDs corresponding to
different $R_{\rm diss}$ (indicated by the vertical grey lines
in Fig. \ref{cmb} with the corresponding numbers); 
all the relevant input parameters for the different SEDs
are reported in Tab. \ref{para}.
The profile of the external radiation and the value of the magnetic
field correspond to what shown in Fig. \ref{cmb}. 
We have also assumed that, increasing $R_{\rm diss}$, the
injected power in relativistic electrons decreases
by a factor 3 increasing $R_{\rm diss}$ by a factor 10.
At the same time, we assumed that the maximum energy
of the injected electrons increases
by a factor 3 for a tenfold increase of $R_{\rm diss}$.
These choices are arbitrary, but reflect the observational evidence
that the bolometric radiative output of jets decreases with distance. 
Furthermore the existence of optical jets, whose emission is due to 
the synchrotron process, ensures that at large jet scales there 
are very energetic electrons.

The first three SEDs (number 1, 2 and 3, corresponding to 
$R_{\rm diss}=10^3$, $10^4$ and $10^5 R_{\rm S}$, respectively) 
have a large Compton dominance,
corresponding to the large ratio between the external radiation 
and the magnetic energy densities. 
Note that only the first uses the BLR photons as the main seeds
for the Compton process (and thus it has the largest peak frequency), 
while the SED 2 and SED 3 use the IR photons as seeds. 
SED 1 is dominating in the $\gamma$--ray band (above $\sim$1 MeV), while
in the far IR to UV bands there is the contribution of the
thermal radiation (accretion disk and IR torus).
Remarkably, the soft X--rays are produced almost equally
by these three jet dissipation sites (see also below).

SED 4 (i.e. $R_{\rm diss}=10^6 R_{\rm S}$)
is the only one where the magnetic energy density dominates.
Correspondingly, the synchrotron flux dominates the bolometric output.
Comparing SED 3 and SED 4, we note that they have total
luminosities that differ by more than a factor 3.
This is due to the incomplete electron cooling, and implies that
not all the power injected in random energy of the electrons can
be radiated in one crossing time.
For the same reason, there is a rather large jump in $\gamma_{\rm cool}$ 
between $R_{\rm diss}=10^5$ and $10^6R_{\rm S}$ 
(see the bottom panel of Fig. \ref{cmb}), corresponding to the 
fast drop of the energy density of the external radiation between
these two distances.

SED 5 ($R_{\rm diss}=10^7 R_{\rm S}$, or 1 kpc)
has again a relatively large Compton dominance,
due to the prevailing of the CMB 
and the starlight (from the galaxy bulge) 
energy densities over the magnetic one.
This SED has almost the largest possible $\gamma_{\rm cool}$
(see Fig. \ref{cmb}) and this is reflected in the calculated
SED, having the largest peak frequency of the high energy component.

SED 6 and 7 ($R_{\rm diss}=10^8$ and $10^9 R_{\rm S}$, or 10 and 100 kpc,
respectively) have a very large Compton dominance, due to the fact
that $U^\prime_{\rm CMB}$ is constant, while $U^\prime_{\rm B}$ 
decreases with distance.
As discussed previously, the constancy of $U^\prime_{\rm CMB}$
makes $\gamma_{\rm cool}$ to decrease with $R_{\rm diss}$,
since the cooling time at which $\gamma_{\rm cool}$ is calculated increases.
This also implies that, despite the decrease of $P^\prime_{\rm i}$,
the SED 5, 6 and 7 have the same bolometric luminosity,
because a larger portion of the electron population can cool
in one $t_{\rm cross}$.

\vskip 0.3 cm
The sum of all SEDs is shown by the black thick line in Fig. \ref{rdiss}.
Note that, in the radio band, the total flux $F_\nu\propto \nu^0$
(or slightly harder), as observed.
At frequencies greater than the radio ones,
the total flux is dominated by SED 1, but the flux originating at relatively 
large scales can be important in some frequency bands, as the 
soft X--ray one, with important consequences on the observed 
variability and the correlation of variability in different bands.
According to Fig. \ref{rdiss} (which is, we re--iterate, only one
possible example, shown for illustration) the correlation 
between the $\gamma$--ray and the IR flux should be tighter
than the correlation between the $\gamma$--ray and the soft X--ray flux,
diluted by the the flux originating at larger scales
(that can vary on longer timescales).
Furthermore, if the soft X--ray flux is produced by external Compton,
the energies of the electron emitting it are relatively modest
(namely $\gamma\sim$1--10)
while, in the $\gamma$--ray band, we see the emission of the 
electrons with the highest energies (which emit, by synchrotron,
at IR--optical frequencies).
The cooling timescales are therefore different, producing
different variability if $t_{\rm cool}$ at low energies is longer 
than the light crossing time.
If not, then we expect the same variability pattern for
emission produced in the same zone.
One example of different variability behaviour is displayed 
by the blazar 3C 454.3, as recently studied by Bonning et 
al. (2008), and has been
interpreted by these authors as a consequence of the different energies 
of the electron contributing in the X--ray and $\gamma$--ray bands.
The contribution to the X--ray flux by other zones of the jet
can be an alternative possibility.

\begin{table*} 
\centering
\begin{tabular}{lllllllllllllll}
\hline
\hline
Fig.   &$R_{\rm diss}$ &$M$ &$R_{\rm BLR}$ &$P^\prime_{\rm i}$ &$L_{\rm d}$  &$B$  &$\Gamma$ &$\theta_{\rm v}$
    &$\gamma_{\rm b}$ &$\gamma_{\rm max}$ &$s_1$  &$s_2$&$z$ &Notes  \\
~[1]      &[2] &[3] &[4] &[5] &[6] &[7] &[8] &[9] &[10] &[11] &[12] &[13] &[14] &[15] \\
\hline   
\ref{pairs}   &6 (20)    &1e9   &1e3 &0.1    &100 (0.67) &200    &10  &3  &1e3 &1e4   &0 &2.5   &3   &Pairs--no pairs    \\
\ref{rpairs}  &3 (10)    &1e9   &1e3 &0.1    &100 (0.67) &1.6e3  &1.8 &3  &1e3 &1e4   &0 &2.5   &3   &Pairs--no pairs   \\
\ref{rpairs}  &30 (100)  &1e9   &1e3 &0.1    &100 (0.67) &120    &5.8 &3  &1e3 &1e4   &0 &2.5   &3   &Pairs--no pairs   \\
\ref{rpairs}  &300 (1e3) &1e9   &1e3 &0.1    &100 (0.67) &5.6    &10  &3  &1e3 &1e4   &0 &2.5   &3   &Pairs--no pairs   \\
\hline
\ref{rdiss} &3e2 (1e3) &1e9  &1e3 &7.3e--2 &100 (0.67) &3.6      &15 &3  &100 &1e4   &1 &2.5   &3   &$R_{\rm diss}$ seq.  \\
\ref{rdiss} &3e3 (1e4) &1e9  &1e3 &2.4e--2 &100 (0.67) &0.36     &15 &3  &100 &3e4   &1 &2.5   &3   &$R_{\rm diss}$ seq.  \\
\ref{rdiss} &3e4 (1e5) &1e9  &1e3 &8.1e--3 &100 (0.67) &3.6e--2  &15 &3  &100 &9e4   &1 &2.5   &3   &$R_{\rm diss}$ seq.  \\
\ref{rdiss} &3e5 (1e6) &1e9  &1e3 &2.7e--3 &100 (0.67) &3.6e--3  &15 &3  &100 &2.7e5 &1 &2.5   &3   &$R_{\rm diss}$ seq.  \\
\ref{rdiss} &3e6 (1e7) &1e9  &1e3 &9e--4   &100 (0.67) &3.6e--4  &15 &3  &100 &8.1e5 &1 &2.5   &3   &$R_{\rm diss}$ seq.  \\
\ref{rdiss} &3e7 (1e8) &1e9  &1e3 &3e--4   &100 (0.67) &3.6e--5  &15 &3  &100 &2.4e6 &1 &2.5   &3   &$R_{\rm diss}$ seq.  \\
\ref{rdiss} &3e8 (1e9) &1e9  &1e3 &1e--4   &100 (0.67) &3.6e--6  &15 &3  &100 &7.3e6 &1 &2.5   &3   &$R_{\rm diss}$ seq.  \\
\hline
\ref{m}     &4.5 (500) &3e7  &55  &3e--3 &0.3 (0.067)   &13.3 &15 &3   &100  &1e4   &1  &2.5   &3 &$M$ seq.        \\
\ref{m}     &15  (500) &1e8  &100 &0.01  &1   (0.067)   &7.3  &15 &3   &100  &1e4   &1  &2.5   &3 &$M$ seq.        \\
\ref{m}     &45  (500) &3e8  &173 &0.03  &3   (0.067)   &4.2  &15 &3   &100  &1e4   &1  &2.5   &3 &$M$ seq.        \\
\ref{m}     &150 (500) &1e9  &317 &0.1   &10  (0.067)   &2.3  &15 &3   &100  &1e4   &1  &2.5   &3 &$M$ seq.        \\
\ref{m}     &450 (500) &3e9  &549 &0.3   &30  (0.067)   &1.3  &15 &3   &100  &1e4   &1  &2.5   &3 &$M$ seq.        \\
\hline
\ref{1253}: 3C 279 &114 (380) &1e9 &173 &0.06  &3 (0.02) &3.1  &11.3 &3 &250 &4e3   &1  &2.2   &0.536 &high EGRET \\
\ref{1253}: 3C 279 &300 (1e3) &1e9 &173 &0.04  &3 (0.02) &0.42 &16   &3 &3e3 &3e5   &0  &2.7   &0.536 &TeV \\
\ref{1253}: 3C 279 &72  (240) &1e9 &173 &9e--3 &3 (0.02) &7.7  &8.9  &3 &160 &1.2e3 &0  &2.7   &0.536 &low EGRET \\
\hline
\ref{1428}: 1428 &680   (1.5e3) &1.5e9 &1.2e3 &0.1   &135 (0.6) &1.2  &15   &2.5 &30  &3e3 &0 &2.4   &4.72 &{\it Beppo}SAX  \\
\ref{1428}: 1428 &6.8e3 (1.5e4) &1.5e9 &1.2e3 &0.1   &135 (0.6) &0.12 &15   &2.5 &30  &3e3 &0 &2.4   &4.72 &$R_{\rm diss}\times 10$ \\
\ref{1428}: 1428 &68    (150)   &1.5e9 &1.2e3 &0.1   &135 (0.6) &44   &7.1  &2.5 &30  &3e3 &0 &2.4   &4.72 &$R_{\rm diss}\times 0.1$ \\
\hline
\ref{2149}: 2149 &960   (800)   &4e9   &1.2e3 &0.1   &150 (0.25) &1.7  &12  &3   &10  &1e3 &--1 &2.6 &2.345 &{\it Swift} data \\
\ref{2149}: 2149 &9.6e3 (8e3)   &4e9   &1.2e3 &0.1   &150 (0.25) &0.17 &12  &3   &10  &1e3 &--1 &2.6 &2.345 &$R_{\rm diss}\times 10$ \\
\ref{2149}: 2149 &96    (80)    &4e9   &1.2e3 &0.1   &150 (0.25) &43   &5.2 &3   &10  &1e3 &--1 &2.6 &2.345 &$R_{\rm diss}\times 0.1$ \\
\hline 
\ref{sequence}   &810   (900)   &3e9   &1.7e3 &0.4   &302 (0.67) &2.4  &13  &3   &100 &3e3   &0   &2.5      &---   &Blazar seq. \\
\ref{sequence}   &810   (900)   &3e9   &636   &0.04  &41 (0.09)  &0.9  &13  &3   &200 &3e4   &0   &2.5      &---   &Blazar seq. \\
\ref{sequence}   &120   (400)   &1e9   &387   &4e-3  &15 (0.1)   &5.5  &11.5 &3  &200 &1.5e4 &0   &2.5      &---   &Blazar seq. \\
\ref{sequence}   &210   (700)   &1e9   &---   &1e-3  &---        &0.2  &15  &3   &600 &2e5   &0   &2.5      &---   &Blazar seq. \\
\ref{sequence}   &210   (700)   &1e9   &---   &5e-4  &---        &0.1  &15  &3   &3e3 &7e5   &0   &2.5      &---   &Blazar seq. \\
\hline 
\end{tabular}
\vskip 0.4 true cm
\caption{List of parameters used to construct the SED shown in Figg. 6--13.
Col. [1]: figure number where the model is shown;
Col. [2]: dissipation radius in units of $10^{15}$ cm and (in parenthesis) in units of $R_{\rm S}$;
Col. [3]: black hole mass in solar masses;
Col. [4]: size of the BLR in units of $10^{15}$ cm;
Col. [5]: power injected in the blob calculated in the comoving frame, in units of $10^{45}$ erg s$^{-1}$; 
Col. [6]: accretion disk luminosity in units of $10^{45}$ erg s$^{-1}$ and
        (in parenthesis) in units of $L_{\rm Edd}$;
Col. [7]: magnetic field in Gauss;
Col. [8]: bulk Lorentz factor at $R_{\rm diss}$;
Col. [9]: viewing angle in degrees;
Col. [10] and [11]: break and maximum random Lorentz factors of the injected electrons;
Col. [12] and [13]: slopes of the injected electron distribution [$Q(\gamma)$] below and above $\gamma_{\rm b}$;
Col. [14]: redshift;
Col. [15]: some notes.
For all cases the X--ray corona luminosity $L_X=0.3 L_{\rm d}$.
Its spectral shape is assumed to be $\propto \nu^{-1} \exp(-h\nu/150~{\rm keV})$.
}
\label{para}
\end{table*}

\section{Changing the black hole mass}

\begin{figure}
\vskip -0.5cm
\psfig{figure=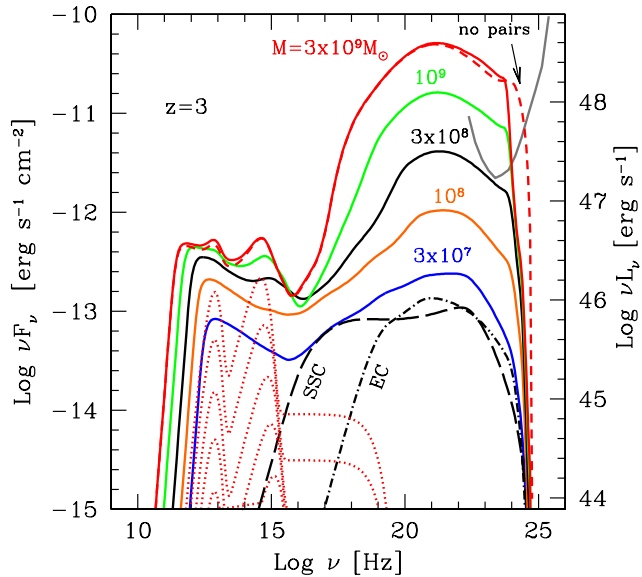,width=9cm,height=9cm}
\vskip -0.5cm
\caption{
The observed SED for different black hole masses
(from $3\times 10^7$ to $3\times 10^9M_\odot$, as labelled)
assuming that the dissipation takes place at 
500$R_{\rm S}$, that the accretion disk luminosity is 
$L_{\rm d} =0.067 L_{\rm Edd}$, and that the
power injected in the jet dissipation region scales as 
$P^\prime_{\rm i} =10^{44}M_9$ erg s$^{-1}$.
The bulk Lorentz factor is kept fixed at $\Gamma=15$.
The particles are always injected between $\gamma_1=10^2$ and 
$\gamma_2=10^4$.
The grey line is the 5$\sigma$ detection sensitivity of Fermi, 
after 1 year of operation.
For the $M=3\times 10^7M_\odot$ case we show, besides the total spectrum,
the SSC and the EC components separately. 
This illustrates the importance of the SSC process for
low values of the black hole masses. 
The dashed line (for the case with $M=3\times 10^9 M_\odot$)
shows the spectrum neglecting photon--photon absorption and pair
reprocessing.
The received flux is calculated assuming that all sources are 
at $z=3$.
}
\label{m}
\end{figure}

It is instructive to study the SED produced by the jet
in FSRQs of different black hole masses, scaling the relevant
quantities with the \sc\ radius and the Eddington luminosity.
To this aim we show in Fig. \ref{m}
a sequence of SED with the black hole mass ranging from 
$3\times 10^7$ to $3\times 10^9 M_\odot$.
We assume that $R_{\rm diss}$ is always at 500 \sc\ radii and
that $L_{\rm d}=0.067 L_{\rm Edd}$, with a corona with
an X--ray luminosity equal to one third that of the disk.
We assume that the power in relativistic electrons injected into
the dissipation region of the jet scales with the black hole mass as
$P^\prime_{\rm i}=10^{44}M_9$ erg s$^{-1}$. 
Electrons are injected
between $\gamma_1=10^2$ and $\gamma_2=10^4$ in all cases.
The bulk Lorentz factor is $\Gamma=15$ for all cases and the redshift is $z=3$.
At the assumed $R_{\rm diss}$ the jet has already reached its
maximum $\Gamma$, and at these distances the Poynting flux 
is assumed to scale as 
$P_{\rm B}=6.7\times 10^{-3} L_{\rm Edd}$.
We then have 
\begin{equation}
U_{\rm B} \, \propto  { L_{\rm Edd} 
\over \Gamma^2 R^2_{\rm diss} } \,
\propto \, {M \over M^2 } \, \propto {1\over M}
\end{equation}
which follows from the assumption of $R_{\rm diss}/R_{\rm S}=$const 
and $\Gamma=$const.
For all our cases, the dissipation occurs within the BLR, which yields a constant
radiation energy density $U^\prime_{\rm BLR}$ since $\Gamma$ is the same.
As a consequence, the ratio 
$L_{\rm EC}/L_{\rm syn} \sim U^\prime_{\rm BLR}/U_{\rm B} \propto M$.
This is the reason of the increasing dominance of the inverse Compton 
emission increasing the black hole mass.
This implies that blazars with large black hole masses
should preferentially be more Compton dominated, and therefore 
more easily detected by the Fermi satellite. 
In fact, in Fig. \ref{m} one can see the 5$\sigma$ sensitivity
of Fermi for 1 year of operation (grey line), suggesting that, at high redshifts,
the detected blazars will preferentially have large black hole masses.

The importance of the EC relative to the SSC emission increases 
with the black hole mass, hardening the X--ray spectral shape.
The SSC and EC components are shown separately
for the SED corresponding to $M=3\times 10^7M_\odot$,
to illustrate the importance of the SSC flux.
For larger masses the SSC components becomes
relatively less important than the EC one.
For the SED with $M=3\times 10^9M_\odot$ we show the effects
of neglecting the $\gamma$--$\gamma$ absorption and the
consequent reprocessing (dashed line).
One can see that the primary (i.e. neglecting pairs)
spectrum as a rather sharp
cut--off due to the Klein--Nishina limit given by
Eq. \ref{kn}.

\section{Some illustrative examples}
\begin{figure}
\vskip -0.5cm
\psfig{figure=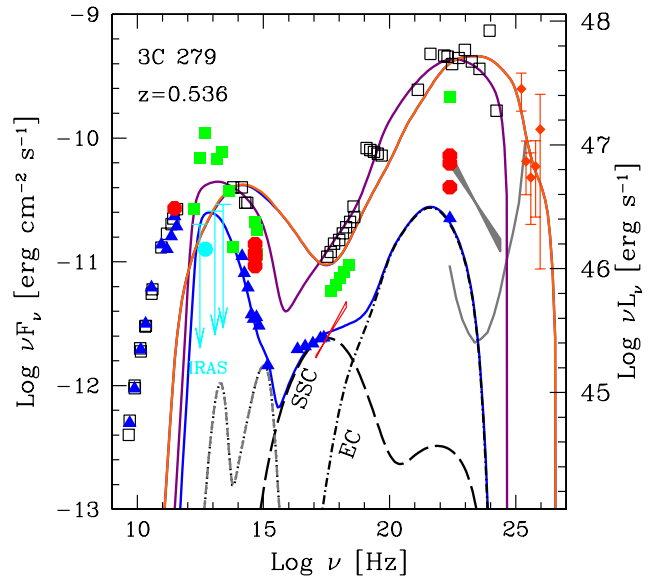,width=9cm,height=9cm}
\vskip -0.5cm
\caption{
SED of 3C 279 in different states together with the corresponding 
models, whose input parameters are listed in Tab. \ref{para}.
See Ballo et al. (2002) and references therein for the sources
of data points.
Note that the high energy data--points (above 100 GeV) have been
de--absorbed according to the Primack, Bullock \& Somerville (2005) 
model used in Tavecchio \& Mazin (2009).
According to Eq. \ref{kn2}, the flux at these energies can
be produced at relatively large distances from the black hole, beyond
the BLR. In this case the main contribution to the seed photons
is coming from the IR torus (solid light grey line, 
orange in the electronic version).
In the low state (blue triangles) the SSC process
is dominating the 2--10 keV X--ray flux.
For this state only we show the SSC (long dashed line) and the EC 
flux (dot--dashed line) separately.
}
\label{1253}
\end{figure}

\subsection{TeV FSRQs: the case of 3C 279}

3C 279 has been recently detected in the TeV band (Albert et al. 2008),
although its redshift, $z=0.536$, implies a strong absorption
of high energy $\gamma$--rays by the IR cosmic background.
This demonstrates that also powerful blazars emit at large
energies, up to the TeV band, even if the peak of their
high energy hump  may lye in the MeV--GeV band.
Besides the consequences that this result has on the 
cosmic background, discussed in Sitarek \& Bednarek (2008); 
Tavecchio \& Mazin (2009); Liu, Bai \& Ma (2008),
we would like to discuss here another consequence,
which concerns the primary spectrum of the source.

We have discussed previously that if the bulk of the
inverse Compton spectrum uses BLR photons as seeds, then 
we expect a steepening of the intrinsic
spectrum at $\sim 15/(1+z)\sim 10$ GeV due to Klein--Nishina effects.
This is shown in Fig. \ref{1253} as the darker line passing through 
an archival EGRET spectrum of the source (squares).
For this model, in fact, the dissipation region is within the BLR,
which is then giving most of the seed photons used for the scattering
(see Tab. \ref{para} for the all the input parameters).
The rather abrupt cut--off seen at $\sim 2.5\times 10^{24}$ Hz $\sim 10$ GeV 
is due this effect, and not to $\gamma$--$\gamma$ internal absorption,
which, in this particular example, is unnoticeable.
To produce photons of larger energies and evade the Klein--Nishina limit 
we are forced to locate $R_{\rm diss}$ beyond the BLR, and then
use the IR photons produced by the torus. 
This is what the model does (lighter grey line, orange in 
the electronic version). 

Note that the presence of a very high energy component in 3C 279
was predicted (before detection) by B{\l}azejowski et al. (2000), who
included the presence of a IR emitting torus (see their Fig. 4).
The main difference of our modelling with respect to B{\l}azejowski et al. 
(2000) is in the assumed temperature of the torus, assumed to be larger 
in that paper, implying a smaller size and an enhanced $U^\prime_{\rm IR}$. 

If the jet of 3C 279 is ``canonical" in the sense described in this paper,
then a larger $R_{\rm diss}$ means a smaller magnetic field, and a larger
bulk Lorentz factor (if it has not yet reached its maximum value).
Then, according to these ideas, we show the entire modelled SED to compare it
with the ``high EGRET state" (squares) and the ``low EGRET state" (triangles).
For the latter model we have assumed a smaller $R_{\rm diss}$ 
and $P^\prime_{\rm i}$.
This implies a small bulk Lorentz factor (hence a decreased importance
of the external seed photons) and a large magnetic field, resulting in
a SED of equal synchrotron and inverse Compton power.

3C 279 is one of the best studied $\gamma$--ray blazars, partly
because it was very active during the observations of EGRET.
It should not be taken, however, as the prototypical high power FSRQ,
since its disk emission is very modest, as also directly
suggested by its SED in the low state (Pian et al. 1999;
see the triangles in Fig. \ref{1253}, see Ballo et al. 2002 
and references therein for the data): for our models we have assumed 
a black hole mass $M=10^9M_\odot$ and $L_{\rm d}=0.02 L_{\rm Edd}$. 
It is therefore instructive to
show example of more powerful blazars, with accretion disk
emitting close to the Eddington limit.

\begin{figure}
\vskip -0.5cm
\psfig{figure=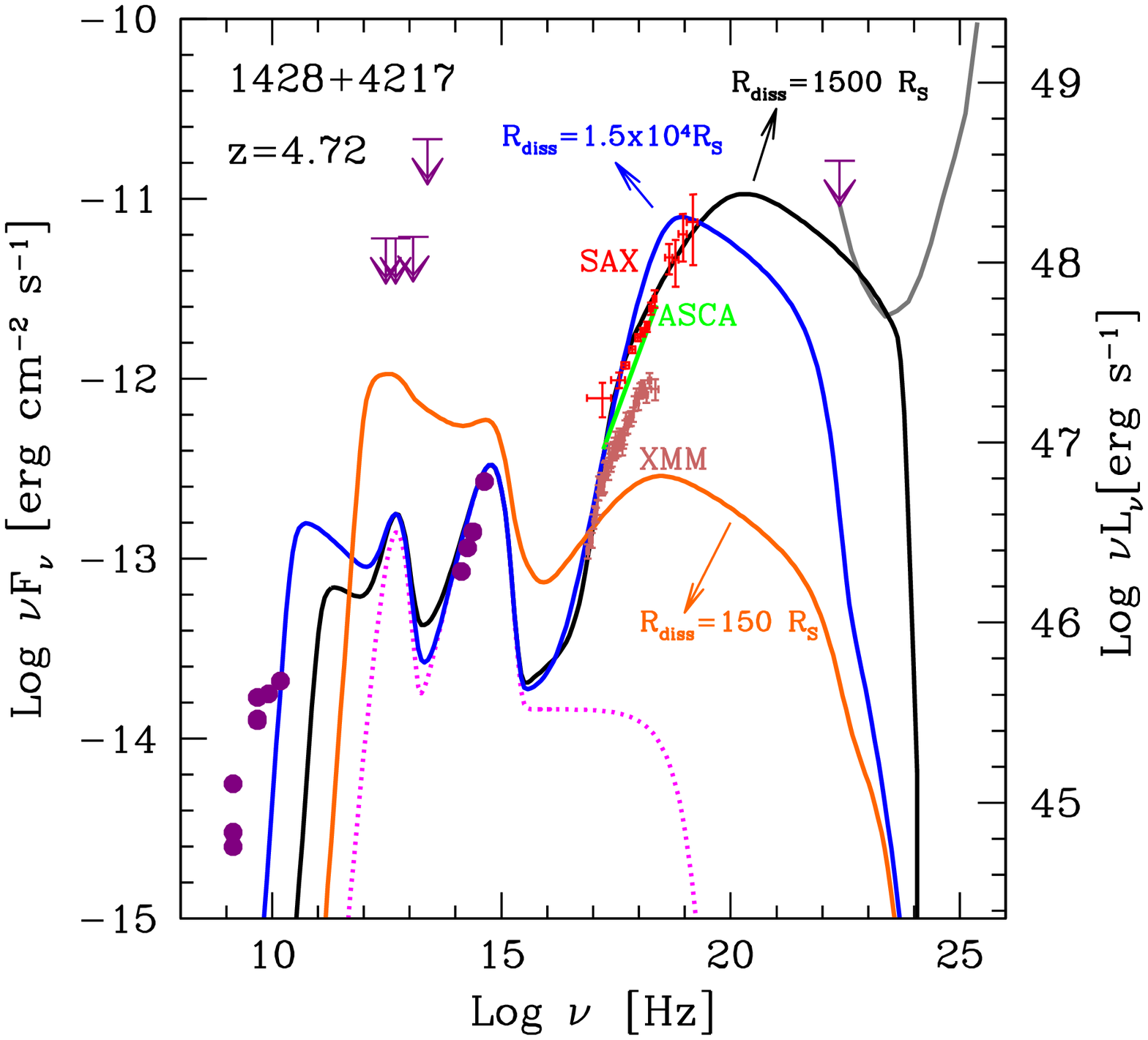,width=9cm,height=9cm}
\vskip -0.5cm
\psfig{figure=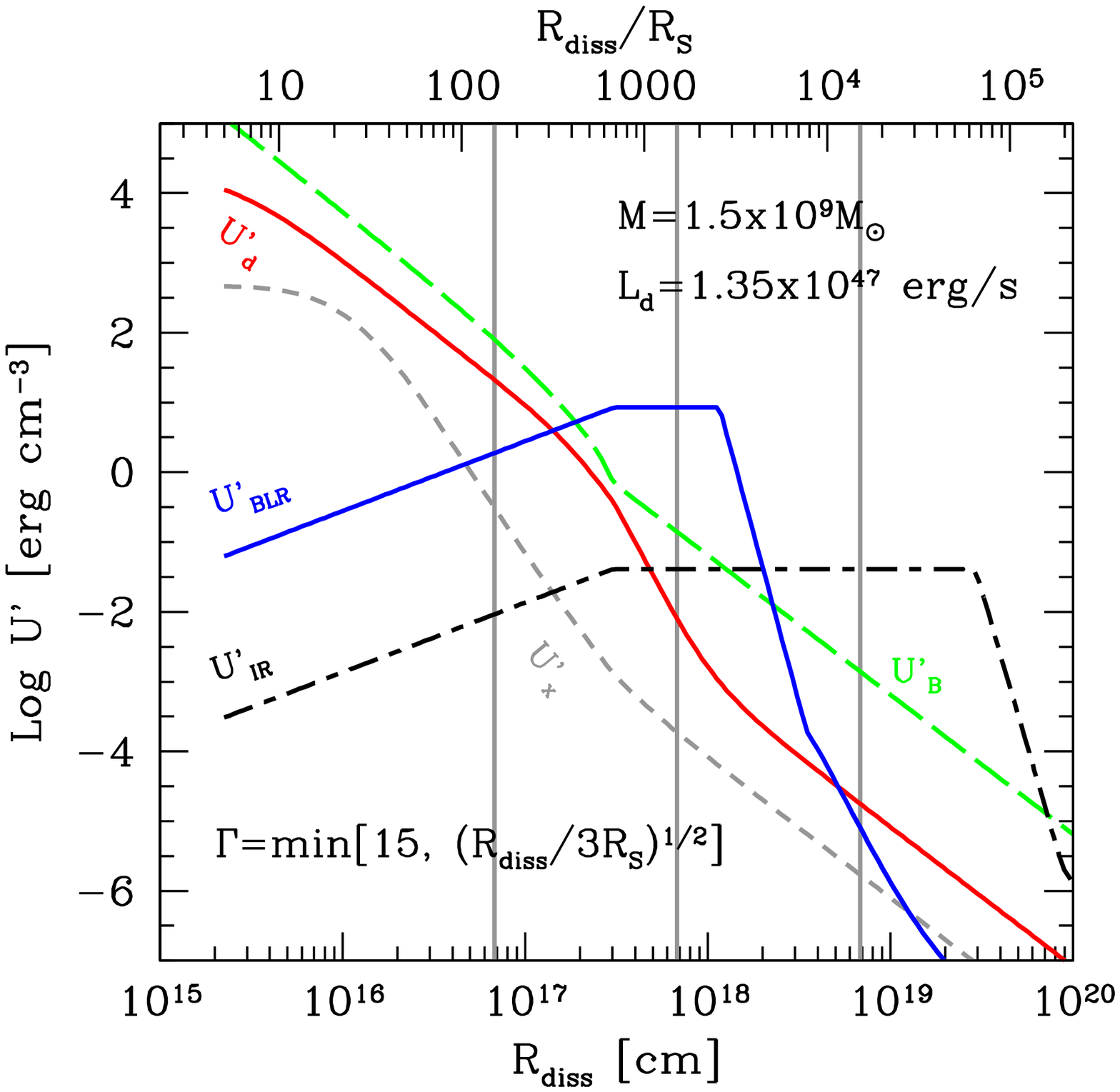,width=8.5cm,height=7cm}
\vskip -0.5cm
\caption{
Top panel: the SED of one of the most distant blazars,
together to 3 different models with different $R_{\rm diss}$ 
(see Tab. \ref{para} for the set of parameters) to illustrate 
possible different states of the source.
See Fabian et al. (2001) and references therein for the sources of data, 
and Celotti et al. (2007) for further discussion about this source.
}
\label{1428}
\end{figure}

\begin{figure}
\vskip -0.5cm
\psfig{figure=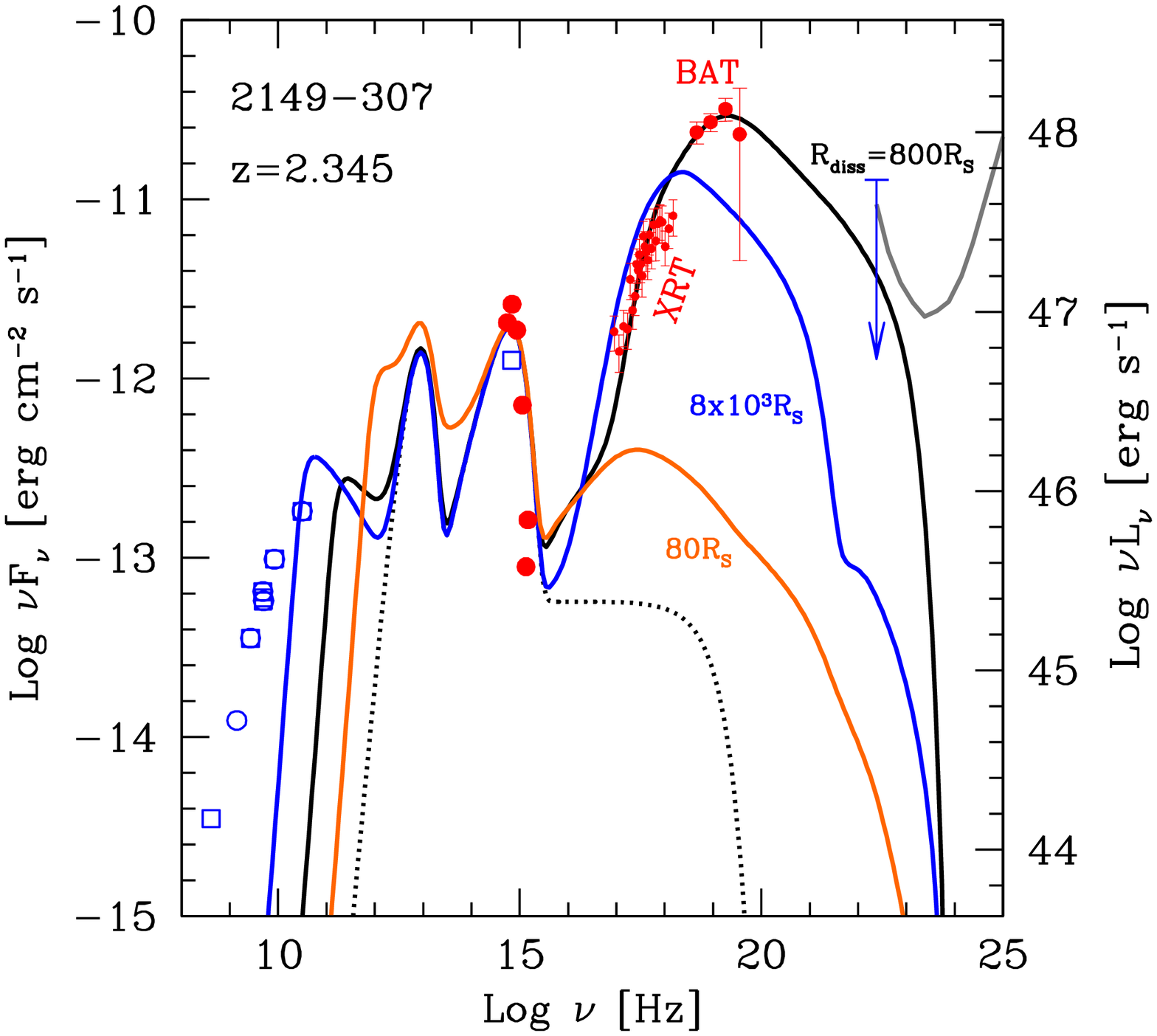,width=9cm,height=9cm}
\vskip -0.5cm
\psfig{figure=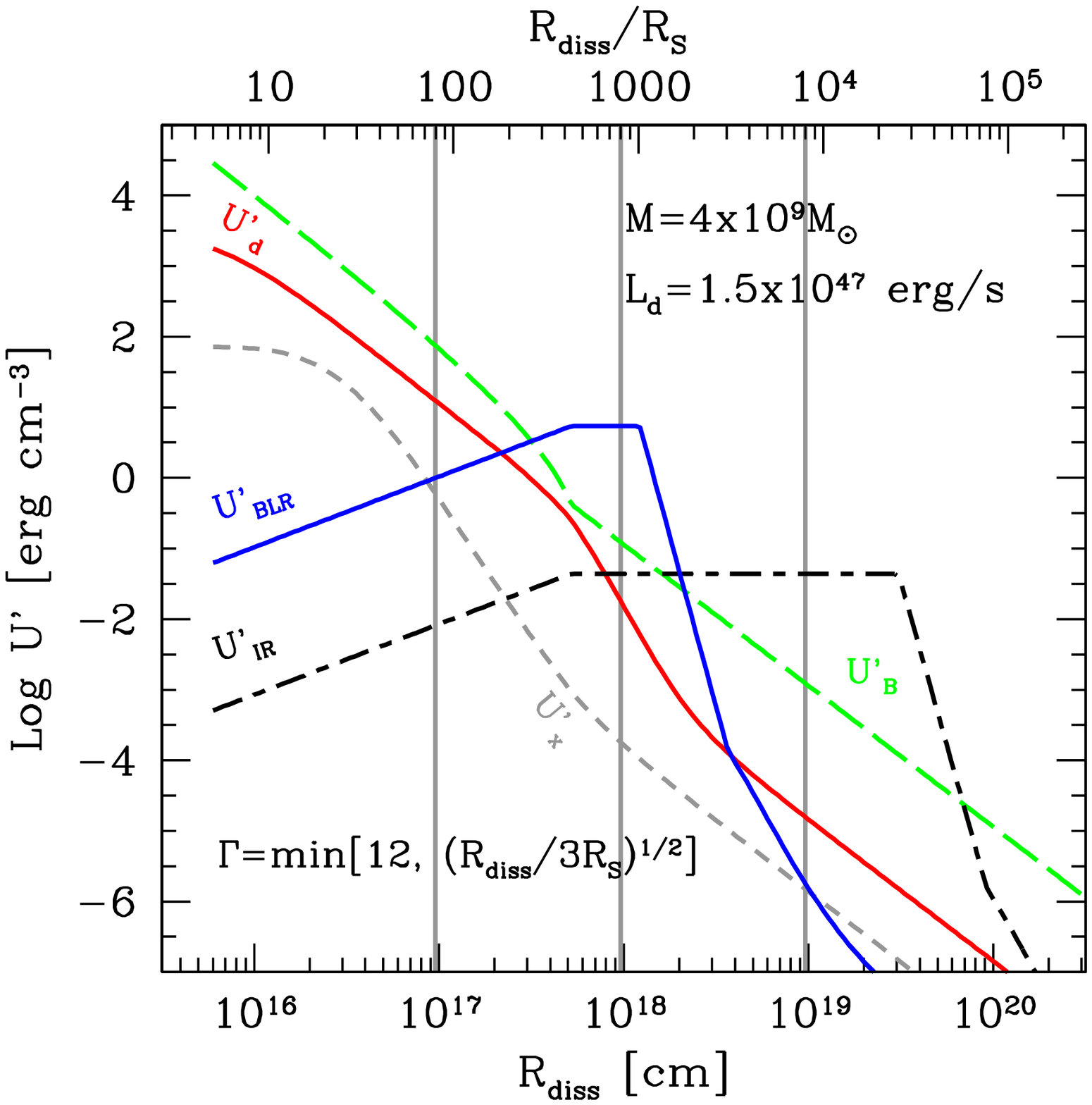,width=8.5cm,height=7cm}
\vskip -0.5cm
\caption{
Top panel: the SED PKS 2149--307 together to 3 different 
models (see Tab. \ref{para} for the set of parameters) 
to illustrate possible different states of the source.
These relatively distant, close to Eddington, high black hole 
mass blazars should be at the extreme of the blazar sequence, 
showing a high energy peak in the 100 keV--1 MeV energy band.
Note that if the dissipation takes place very close to the black hole,
when the jet is still accelerating and with a strong magnetic field,
the resulting spectrum becomes unconspicuous at high energies, even if 
the intrinsic dissipated power is the same of the higher states.
}
\label{2149}
\end{figure}

\subsection{MeV blazars at large redshifts}

GB 1428+4217, at $z=4.73$, is the second most distant blazar known
(the most distant is Q0906--6930, with $z=5.47$).
Its SED is shown in the top panel of Fig. \ref{1428}, together with
the fitting model (black line).
The figure shows that this source has a prominent UV--bump,
with a luminosity around $10^{47}$ erg s$^{-1}$ (see also Tab. \ref{para}).

According to our model, the peak of the high energy hump of the SED
during the {\it Beppo}SAX observations (Fabian et al. 2001) 
is predicted to lye at $\sim 1$ MeV, locating this source at 
one extreme of the blazar sequence.
The model assumes a black hole mass $M=1.5\times 10^9 M_\odot$, an
accretion disk emitting at $\sim 2/3$ of the Eddington ratio (the other third
is emitted by the X--ray corona).
The dissipation region is within the BLR, at 1500 \sc\ radii from the centre.
The fit requires $B=1.2$ G, corresponding to $P_{\rm B}=5.4\times10^{45}$
erg s$^{-1}$, and $L_{\rm d}=1.35\times 10^{47}$ erg s$^{-1}$.
The bottom panel of the same figures shows the radiation and magnetic 
energy density profiles for the parameters used to model this blazars.
The mid grey vertical line corresponds to the used value of $R_{\rm diss}$.
This figure shows that $U^\prime_{\rm BLR}$ is about two orders of magnitude
larger than $U^\prime_{\rm B}$, corresponding to the ratio of the external 
Compton to synchrotron luminosities.

For illustration purposes, we also show the SED corresponding to increase
(decrease) $R_{\rm diss}$ tenfold (see the vertical grey lines in the bottom
panel, and the corresponding SED in the top panel).
Shifting $R_{\rm diss}$ to $1.5\times 10^4 R_{\rm S}$ the high energy peak
moves to lower frequencies, as the relevant seed photons are softer,
and at the same time the X--ray spectrum hardens.
This is obtained even if the particle injection function is unchanged,
and is due to the incomplete cooling occurring for the larger $R_{\rm diss}$ case.
Since $\gamma_{\rm cool}$ shifts from 1 to $\sim$34, in the 
X--ray band we see the emission from the uncooled electron populations, 
that retains the original, ``injection", slope $s=0$.
Note that the high energy peak is now slightly below 100 keV, and this would
make the source even more extreme in terms of the blazar sequence.
The EC to synchrotron luminosity ratio is still $\sim 100$, corresponding
now to the ratio between $U^\prime_{\rm IR}$ and $U^\prime_{\rm B}$,
as can be seen in the bottom panel.

Instead, if $R_{\rm diss}$ is at 150 \sc\ radii, the SED
changes more dramatically. 
Since in this case the bulk Lorentz factor is smaller and the magnetic field
larger, the inverse Compton and synchrotron powers become comparable
(see the bottom panel, showing that for this $R_{\rm diss}$ we have
$U^\prime_{\rm B}\sim U^\prime_{\rm d}$).
The X--ray spectrum brightens and softens considerably, in the IR--UV
band the flux increases (because of the increased magnetic field)
even if the bolometric observed luminosity decreases because of the decreased 
Doppler boosting.
The fact that we did not see this kind of SED in GB 1428+4217 suggests that 
this state rarely occurs.
This is particularly true considering that this blazars 
was not discovered because it was particularly bright in hard X--rays or 
in the $\gamma$--ray band, so there was no bias against a soft X--ray spectrum.

\vskip 0.3 cm

We lastly consider PKS 2149--307 at $z=2.345$, as observed by the XRT and BAT 
instruments onboard {\it Swift} during its first 9 months of observations 
(Sambruna et al. 2007).
Fig. \ref{2149} shows its SED (top panel) and the profiles of the radiation
and magnetic energy densities for the considered models (bottom panels).
As done for GB 1428+4217, we show what we considered the best fitting model
(black line in the top panel) corresponding to $R_{\rm diss}=9.6\times 10^{17}$ cm,
(corresponding to 800 $R_{\rm S}$ for the assumed black hole mass of 
$M=4\times 10^9 M_\odot$), and also the models corresponding to a tenfold 
increase (decrease) of $R_{\rm diss}$ (light grey lines in the top panel and 
vertical lines in the bottom panel).
The BAT data have large error bars, precluding the possibility to firmly claim that
its high energy peak is within the BAT energy range (namely, at $\sim 100$ keV),
but this possibility is indeed suggested by the present data.
The SED of this source can be explained by a set of parameters
similar to the ones chosen for GB 1428+4217, and similar considerations apply.
Increasing $R_{\rm diss}$ tenfold makes the IR radiation from the torus to dominate,
leaving almost unchanged the ratio between the radiation and the magnetic energy 
densities (and thus the corresponding EC to synchrotron luminosity ratio).
Again a decrease in $R_{\rm diss}$ by a factor 10 has a dramatic impact on the
predicted SED, dominated in this case by the synchrotron luminosity (see
also the bottom panel).
However, having only data up to the medium--soft X--ray energy range,
it would be almost impossible to tell what is the real appearance of the
bolometric SED of the source, as almost all the changes occur above
10 keV.

\begin{figure} 
\vskip -0.5cm
\hskip -0.5 cm
\psfig{figure=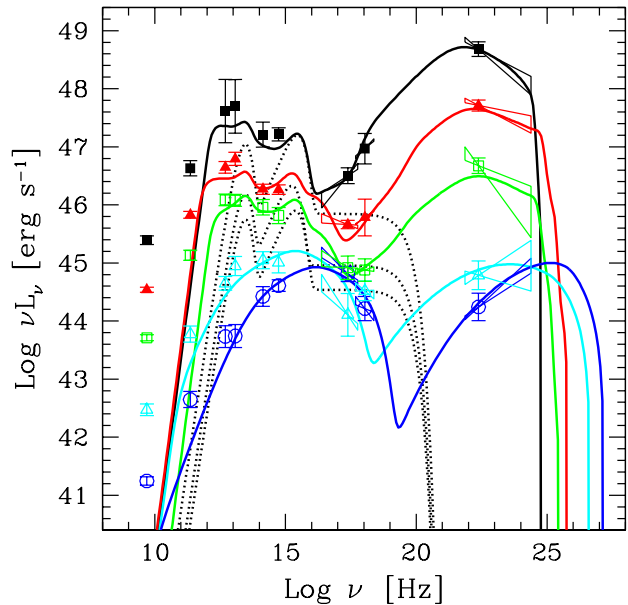,width=9.5cm,height=9.5cm}
\vskip -0.5cm
\caption{
The blazar sequence (Fossati et al. 1998; Donato et al. 2001)
interpreted in the framework of our canonical jet
scenario. Parameters are listed in Tab. \ref{para}.
The SED changes according to changing the accretion rate
and the power of the jet, and assuming that below
some critical accretion rate the accretion regime
changes regime, becoming very radiatively inefficient.
In out case, this occurs for the two least powerful SED,
that should corresponds to low power, line--less BL Lacs.
}
\label{sequence}
\end{figure}

\section{Canonical jets and the blazar sequence}

We can ask if our ``canonical" jet can reproduce the
phenomenological blazar sequence as proposed by Fossati et al. (1998)
(see also Donato et al. 2001 and Ghisellini et al. 1998).
This sequence was constructed by taking flux limited samples
of blazars (in the radio and X--ray bands),
dividing the sources in radio luminosity bins, and
averaging the flux of the sources in each (radio luminosity)
bin at selected frequencies. 
The resulting data are shown in Fig. \ref{sequence}.
As discussed in Ghisellini \& Tavecchio (2008)
and in Maraschi et al. (2008), this sequence does
not pretend to describe ``the average blazar",
since it likely represents the most beamed sources,
successfully entering in the flux limited sample 
they belong to.
Nevertheless it is appropriate to ask if our ``canonical" jet can 
reproduce this sequence without strong modifications of the 
basic assumptions.
Fig. \ref{sequence} then shows the SED resulting from
our modelling, whose parameters are listed in Tab. \ref{para}.
As can be seen the agreement is quite good, and it is achieved
by assuming (as in Ghisellini \& Tavecchio 2008), that the
accretion disk becomes radiatively inefficient below
some critical accretion rate (thus around a critical
luminosity in units of the Eddington one, that we take
as a few $\times10^{-3}$).
The first three, more powerful, SED correspond to sources
with a standard disk, while for the two SED at lower luminosities
we have ``switched--off" all external radiation.
The black hole mass is $3\times 10^9M_\odot$ for the 
two most powerful SED, and is $10^9M_\odot$ for the
other three.
The luminosity, in the comoving frame, injected in relativistic electrons
is monotonically decreasing, as the value of the magnetic field.
The dissipation radius $R_{\rm diss}$, in units of the \sc\ radius,
changes slightly, but less than a factor 2.5.
The energy of the injected particles need to increase as the 
cooling decreases.
The decreased cooling also makes $\gamma_{\rm cool}$ to increase.
These effects are very important for the two ``BL Lac"
SED, and much less for the other three most luminous SEDs.

We can conclude that the blazar sequence, as illustrated in 
Fig. \ref{sequence}, can be interpreted as a sequence of jet powers, 
in sources having more or less the same (large) black hole mass.
The jet power correlates with the accretion disk luminosity
and below some critical value the disk changes
accretion regime, becoming radiatively inefficient and
unable to photo--ionise the broad line clouds. 
A intriguing manifestation of this behavior is shown
by the bright blazars recently detected by the {\it Fermi} satellite
(Abdo et al. 2009) discussed in Ghisellini, Maraschi \& Tavecchio 
(2009), and mirrors what has been suggested to occur for the 
FRI--FRII radio--galaxy divide (Ghisellini \& Celotti 2001).

\section{Caveats}
\label{caveats}

Our aim was to describe the main properties of the produced
spectrum in a ``canonical" jet.
Specific sources can of course deviate somewhat by our description. 
For instance:

\begin{itemize}
\item
Assuming $R_{\rm BLR}\propto L_{\rm d}^{1/2}$,
may be approximately obeyed on average,
but specific sources might behave differently.
The assumed jet geometric profile 
and the assumption that the jet, at its start, 
is magnetically dominated could be two oversimplifications.
We hope to test 
these assumptions
by direct observations, since the combined efforts 
of the {\it Fermi} and {\it Swift} satellites
can give us really simultaneous spectra on a large
frequency band.

\item
We have assumed that the ``dissipation"
mechanism, converting some fraction of the jet power into
radiation, has the form of an injection of primary leptons
lasting for a light crossing time.
With this assumption, we have derived the particle distribution
at this time, which corresponds to the maximum produced flux.
However, the injection can last for a longer or shorter time.
Our framework is appropriate for describing a ``snapshot"
of the SED, and not for a time--dependent analysis.

\item
We assume that the magnetic field in the dissipation region 
is the same as the magnetic field transported by the jet. 
It is not amplified by e.g. shocks, nor it is  reduced
by e.g. reconnection events (see e.g. Giannios, Uzdensky \& Begelman 2009).
Although we can assume any value of the magnetic field
in the dissipation region (i.e. we can fix any value
in the numerical code calculating the SED), we prefer this choice 
for all the examples shown, because of simplicity.

\item
We focused on high power blazars, thought to have ``standard" 
accretion disks and broad line regions, and therefore we did 
not discuss in detail the expected SED from low power 
(and TeV emitting) BL Lacs. 
However, we could simply extend our study to these sources
if we ``switch--off" the external radiation components,
as expected if the accretion disk becomes radiatively inefficient,
making very weak or absent broad emission lines.
In Fig. \ref{sequence} we indeed show two examples of
the expected spectra in this case.

\item
All the shown SED do not take into account the
absorption of $\gamma$--rays due to the IR, optical
and UV background, but only the ``internal" absorption 
caused by the radiation fields existing close to the 
emitting blob.

\item
We have always assumed a relatively small viewing
angle ($\theta_{\rm v}\sim 3^\circ$), of the same order of $1/\Gamma$.
Since $\Gamma$ and $\theta_{\rm v}$ are two separated input parameters,
we could change $\theta_{\rm v}$ (and thus the Doppler factor $\delta$)
for a fixed $\Gamma$. 
However, this would introduce a delicate issue, since the external 
radiation field, in the comoving frame, is not isotropic.
If the emitting blob is inside the BLR, for instance, then
all external photons appear to come from one direction, and
we could apply the formalism of Dermer (1995) to describe
the pattern of the scattered radiation.
Note that, in this case, if $\theta_{\rm v}=1/\Gamma$
(and thus $\delta=\Gamma$) one has the observed luminhosity
as for radiation isotropic in the rest frame.
On the other hand, if the emitting region is beyond the BLR, 
then the arrival directions of the BLR external photons are 
spread, and it is not trivial to reconstruct the exact pattern. 
But in this cases the BLR component is hardly dominant with respect to
the IR radiation from the torus.
We have decided, for simplicity, to use $\theta_{\rm v}\sim 1/\Gamma$,
and to treat the emitted radiation pattern is the same way as for an 
isotropic seed photon distribution.

\item 
We neglected, for simplicity, the possibility of a jet 
composed by a fast spine surrounded by a slower layer.
This structured jet have been proposed for low power
BL Lacs and FR I radio--galaxies (Ghisellini, Tavecchio 
\& Chiaberge 2005), and can ease some problems in explaining the 
SED properties in these sources.
While in high power blazars there is no compelling need, yet,
for such a structure, its inclusion in the present study
would make our description much more complex and model--dependent,
given the freedom to choice the parameters for the layer emission.

\item 
For simplicity, we neglected the possibility that the BLR and 
the associated inter--cloud material
can absorb and re-emit, or scatter, part of the synchrotron radiation
produced by the jet (i.e. the ``mirror model", see Ghisellini \& Madau 1996,
and some criticism by Bednarek 1998; B\"ottcher \& Dermer 1998).

\item
The jet, even if cold (i.e. before dissipation), 
is relativistic, and the carried leptons could scatter 
the external radiation by bulk Comptonization. 
This process, proposed by Begelman \& Sikora (1987) and 
Sikora et al. (1994) and studied
in detail in Celotti, Ghisellini \& Fabian (2007), could produce
a black--body like component in the X--ray band, whose level
depends on how the jet accelerates and on the amount of
leptons carried by the jet.
We have neglected this component for simplicity, although
it can be a very important diagnostic for deriving, at the same
time, the bulk Lorentz factor of the jet and its matter content.

\end{itemize}

\section{Conclusions}

The relatively recent observations of blazars, especially
in the $\gamma$--ray band, have disclosed some of the 
crucial ingredient for our understanding of jets, namely
the bolometric power output and the real shape of 
the produced SED.
Although there is still discussion about the origin
of the high energy emission of blazars, the relatively simple
one--zone leptonic model was rather successful to
interpret the existing data of specific sources.

In this paper, rather than modelling single sources,
we tried to propose a more general description of the emission
produced by high power jets.
For this aim we have first summarised in a simple
way the expected radiation fields as a function of the 
distance from the black hole, and then we have studied 
how the expected SED changes if the main dissipation
occurs at different locations in the jet.
We have done this by assuming rather reasonable prescription for
the accretion disk and its X--ray corona emission, and
using reasonable prescriptions for the radiation fields
corresponding to the broad line region and the infrared torus.
We have also assumed that the jet conserves its Poynting flux
at all distances, thus fixing the profile of the magnetic field.
Since all these prescriptions correspond to rather uncontroversial
assumptions, we believe that they should well characterise the
properties of an average, canonical, large power jet.

The main conclusions of our work are:

\begin{itemize}

\item 
The magnetic energy density is
always smaller than the external radiation energy densities, except
at the very beginning of the jet and at the kpc scale. 
Beyond these scales, the contribution of the cosmic microwave background
becomes dominant.

\item 
The inclusion of the X--ray radiation from the disk corona 
allowed to quantify the effect of pair production and reprocessing
for small $R_{\rm diss}$.
This is found to be severe if the primary spectrum emits
most of its power at energy above the pair production threshold.
This requires that the bulk Lorentz factor is large
even in the very vicinity of the black hole.
In this case 
there is a transfer of power from the $\gamma$--ray to the X--ray band.
Our quantitative analysis confirms earlier, more qualitative, statements
concerning the non--dissipative nature of the inner (i.e. up to a few hundreds
\sc\ radii) jet.

\item 
If instead the jet accelerates gradually (i.e. as $\Gamma\propto R^{1/2}$
up to a maximum value), the magnetic energy density dominates
the cooling at the start of the jet,
resulting in a synchrotron dominated SED,
with a very small fraction of power being emitted above the pair 
production energy threshold.
These SED would hardly be detectable, since the small $\Gamma$
means a much reduced Doppler enhancement of the flux, and thus
these components are easily overtaken by the emission at larger distances
(that have a larger $\Gamma$).
In any case, early dissipation (i.e. small $R_{\rm diss}$) 
is not an efficient process to create electron--positron pairs.

\item 
Besides the inner part of the jet, the ``internal" pair absorption 
(i.e. calculated within the emitting blob) can be particularly important 
also if dissipation takes place close to (but within) the BLR and
close to (but within) $R_{\rm diss}\sim R_{\rm IR}$.
In these cases, however, a small fraction of the 
emitted power can be absorbed, and the reprocessing is modest.

\item
The Klein--Nishina decrease (with energy) of the scattering cross section 
imprints a characteristic steepening in the spectrum above $\sim 10$ GeV
if the dissipation takes place within the BLR, and at $\sim 1$ TeV if
the dissipation occurs beyond the BLR but within $R_{\rm IR}$.

\item 
The above point has important consequences
on the possible use of the high energy (GeV) data of blazars to
study the optical--UV background.
In fact the signature of the $\gamma$--$\gamma$ absorption 
due to this component would be a rather sharp steepening 
of the received spectrum, that would be very similar to the
intrinsic steepening of the primary spectrum due to 
Klein--Nishina effects {\it if most of the dissipation
takes place within the BLR}. In this case also the 
``internal" $\gamma$--$\gamma$ absorption due to BLR photons
will be relevant.
It will be very difficult to disentangle these effects. 
On the other hand, if the observed spectrum continues unbroken above
$15(1+z)$ GeV, we can infer that most of the dissipation
has occurred beyond $R_{\rm BLR}$. 
In this case both Klein--Nishina and ``internal" absorption effects are 
unimportant up to the TeV band.
These therefore would be the best candidates to study the 
opt--UV cosmic backgrounds.

\item
The magnetic field decreases with distance, while
the $U^\prime_{\rm BLR}$ is approximately constant up to $R_{\rm BLR}$,
and $U^\prime_{\rm IR}$ is approximately constant up to $R_{\rm IR}$.
At these distances (i.e. $R_{\rm diss}\sim R_{\rm BLR}$ and
$R_{\rm diss}\sim R_{\rm IR}$) the Compton dominance
is very large.
Existing observations of large power blazars already indicate that 
these blazars are characterised by a weak synchrotron component,
a strong thermal, disk component (unhidden by the synchrotron flux)
and a very large high energy flux. 
These are the sources with the largest Compton dominance,
strongly suggesting that $R_{\rm diss}$ is relatively large
(i.e. close to the $R_{\rm BLR}$ or $R_{\rm IR}$).
This helps to understand why in these sources the soft X--rays
are relatively weak and very hard, demanding a very weak SSC flux:
since the magnetic field is small, not only the synchrotron, but also
the SSC flux is small with respect to the EC radiation.

\item
Concerning the above point, the fact that the high power blazars
we already know of have the mentioned properties does not imply that
they preferentially dissipate at large distances with respect to less
powerful sources, since we may have selected them on the basis of
their hard X--ray or $\gamma$--ray flux (therefore taking advantage of 
the large Compton dominance they have when $R_{\rm diss}$ is large).
In other words: they may dissipate at different $R_{\rm diss}$ at
different times, but we select them only when $R_{\rm diss}$ is large.

\item 
Although most of the dissipation should take place in
one zone of the jet (with $R_{\rm diss}\sim$500--1000 $R_{\rm S}$),
it is very likely that dissipation occurs also at larger
distances, albeit with less power.
The flux originating in these zones can be important in
some frequency bands, as the soft X--ray one, with
important consequences on the observed variability.
Instead, in the $\gamma$--ray energy band (at least when the $\gamma$--ray
dominates the bolometric output) most of the flux should originate
in one zone. 

\item
Blazars with large masses should be characterised by the largest 
inverse Compton to synchrotron ratios, and thus they will be
well represented in $\gamma$--ray all sky surveys, as
performed by the {\it Fermi} satellite.

\item
The most powerful blazars, with the most powerful associated accretion disk,
should have a high energy hump peaking in the 100 keV--1 MeV energy range.
The exact location of the peak frequency depends somewhat on $R_{\rm diss}$,
being smaller if dissipation occurs beyond the BLR. 
These blazars should form the high power extension of the blazar sequence.
They {\it are not} easily detectable by the {\it Fermi} satellite, since their MeV
peak implies relatively small fluxes in the GeV band.
The brightest of them have already been revealed by the serendipitous
survey of the BAT instrument onboard {\it Swift}.

\item
The phenomenological blazar sequence as presented in Fossati et al. (1998) 
can be interpreted as a sequence of ``canonical jets" in sources with
more or less the same  (large) black hole mass, but
with different accretion rates and different jet powers.

\end{itemize}

\section*{Acknowledgments}
We sincerely thank the referee for criticism that helped to
substantially improve the paper. 
We thank fruitful discussions with Laura Maraschi and Annalisa Celotti.
This work was partly financially supported by a 2007 COFIN-MIUR grant.

\end{document}